\documentclass[aps,prx,twocolumn,amsmath,amssymb]{revtex4}

\usepackage[version=3]{mhchem} 
\usepackage[T1]{fontenc}       

\usepackage{graphicx}
\usepackage{graphicx}
\usepackage{Jonasmacros} 
\usepackage{bm}
\usepackage{mathptmx}
\usepackage{epstopdf} 
\usepackage{txfonts}

\usepackage[colorlinks,citecolor=blue,linkcolor=Fuchsia]{hyperref}
\usepackage[usenames,dvipsnames,svgnames]{xcolor}

\begin{document}

\date{\today} 
\title{Charge Transport and Entropy Production Rate in Magnetically Active Molecular Dimer} 

\author{J. D. Vasquez Jaramillo}
\affiliation{Department of Physics and Astronomy, Box 516, 75120, Uppsala University, Uppsala, Sweden}

\author{J. Fransson}
\email{Jonas.Fransson@physics.uu.se}
\affiliation{Department of Physics and Astronomy, Box 516, 75120, Uppsala University, Uppsala, Sweden}


\begin{abstract}
We consider charge and thermal transport properties of magnetically active paramagnetic molecular dimer. Generic properties for both transport quantities are reduced currents in the ferro- and anti-ferromagnetic regimes compared to the paramagnetic and efficient current blockade in the anti-ferromagnetic regime. In contrast, while the charge current is about an order of magnitude larger in the ferromagnetic regime, compared to the anti-ferromagnetic, the thermal current is efficiently blockaded there as well. This disparate behavior of the thermal current is attributed to current resonances in the ferromagnetic regime which counteract the thermal flow. The temperature difference strongly reduces the exchange interaction and tends to destroy the magnetic control of the transport properties. The weakened exchange interaction opens up a possibility to tune the system into thermal rectification, for both the charge and thermal currents.
\begin{center}
\includegraphics[width=9cm]{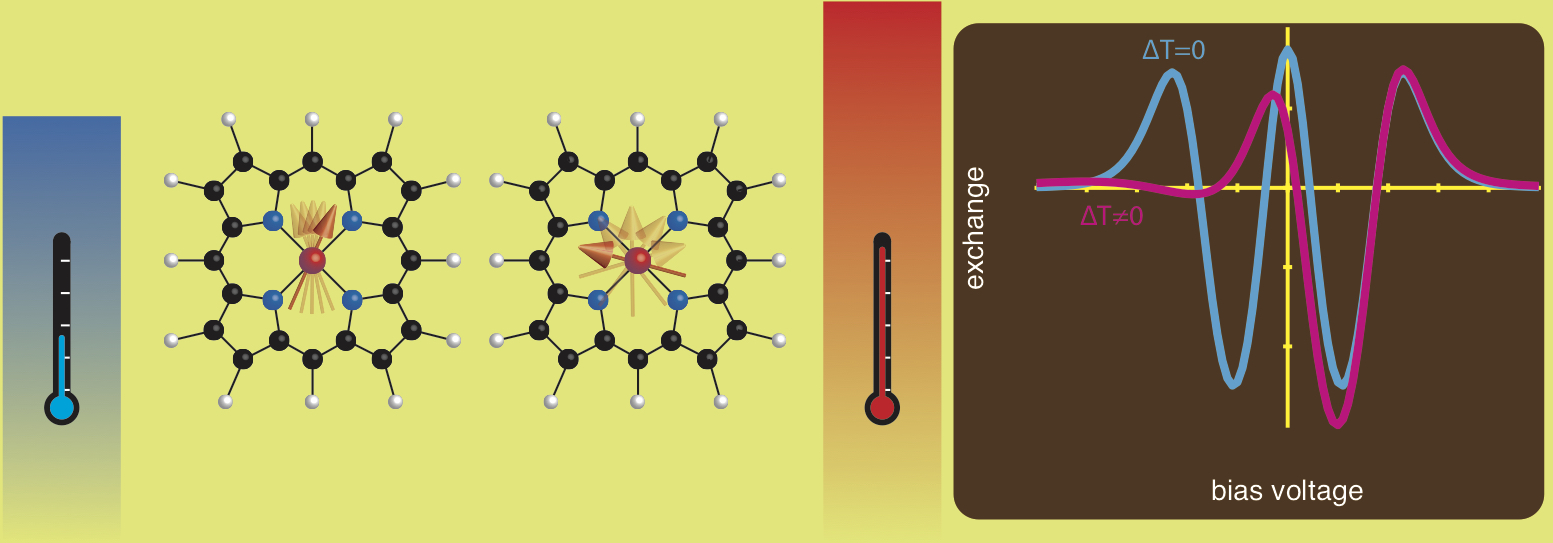}
\end{center}
\end{abstract}

\maketitle

\section{Introduction}
\label{sec-introduction}

Thermal transport properties in molecular junctions can be of electronic origin or mediated through lattice vibrations \cite{Huang1987}. There has been an increasing interest in the study of thermal properties in molecular junctions \cite{Nitzan2007,Pekola2015,Reddy2007,Garrigues2016}, stimulated by experimental observations \cite{Nitzan2007}. Several realizations of tunneling junctions comprising noble metal electrodes and polymers that absorb, emit, and transmit thermal have been reported \cite{Reddy2007,Osorio2010}. The interest is, moreover, driven from the perspective of information science and technology with respect to entropy production rate \cite{Parrondo2015,Pekola2015} and the meaning of the thermodynamics in low dimensional systems \cite{Pekola2015,Esposito2015a,Shastry2016}. This was motivated by the discovery of conducting polymers and solitonic electronic transport mechanisms discovered by Shirakawa, see \cite{Dauxois2006} and references therein. By this discovery polyacetylene became the test bench, bridging the gap between organic and inorganic chemistry regarding electronic transport \cite{Chiang1977,Fincher1978,Schrieffer1980}. Since then, charge transport has been extensively studied theoretically and experimentally in molecular junctions \cite{Rammer1986,Heeger1988,Roth1989,Nakata1992,Meir1992,Bao1996,Parthasarathy1998,Chen2002,Chen2003,Haibo2009,Haug2013}. From the same standpoint, thermal transport studies were conceived and came to be conclusive in the upcoming years \cite{Nitzan2007}. Subsequently, several theoretical studies demonstrated the possibility to conduct both thermal and charge through tunneling junctions \cite{Giazotto2006,Dubi2011,Liu2009,Ludovico2014,PhysRevB.92.235440,Dare2016,Arrachea2014,Zhou2015,Segal2015a}, both in presence and absence of lattice vibrations, although there is no generic framework that successfully is capable of describing the thermodynamic properties of nano junctions \cite{Esposito2015a,Ye2016a,PhysRevB.94.035420}.

Theoretical predictions suggest all electrical control for both reading and writing spin states in molecular dimers \cite{Saygun2016a,Diaz2012}. This prediction is based on the (electronically mediated) indirect exchange interaction between the localized spins which controls the charge transport properties \cite{Saygun2016a,Diaz2012}. Similar effects were reported in Ref. \cite{Osorio2010}, where the spin ground state of a single metal complex is electrically controlled, imposing transition between high ($S=5/2$) and low ($S=1/2$) spin configurations in a three terminal device. 

Here we build on the previous predictions made in Ref. \cite{Fransson2014,Saygun2016a} for dimers of magnetic molecules in which the effective spin-spin interactions are mediated by the properties of the delocalized electrons and extend to thermally induced magnetic and transport properties. In particular we study thermal transport and its response to changes of the magnetic configurations.
Our set-up pertains to, for instance, M-phthalocyanine (MPc), M-porphyrins, where M denotes a transition metal atom \cite{Heinrich2013,Heinrich2013a,Heinrich2015,Brumboiu2016}, e.g., Cr, Mn, Fe, Co, Ni, Cu, and also to bis(phthalocyaninato)R (TPc$_2$) \cite{Urdampilleta2011,Vincent2012}, where R denotes a rare earth element, e.g., Tb. Such molecules can be investigated in, for example, mechanically controlled break-junctions \cite{Liu2014,Vincent2012}, in carbon nanotube assemblies \cite{Urdampilleta2011} and scanning tunneling microscope \cite{Heinrich2013a,Heinrich2015}.

We consider thermal and charge transport as the result of the electrothermal control of the junction. Accordingly, we investigate the charge and thermal conductance with respect to the bias voltage and thermal gradient across the junction. With this background we, furthermore, consider a non-equilibrium analogue to the Seebeck coefficient defined as the ratio between the differential conductances with respect to the thermal gradient and bias voltage, introduced in \cite{Fransson2011}. By the same token, we consider the ratio of the energy differentials with respect to the thermal gradient and the bias voltage. Our predictions and results are based on non-equilibrium Green functions defined on the Keldysh contour.

Throughout this study we consider spin degenerate conditions in that we assume non-magnetic metals in the leads as well as the absence of externally applied magnetic fields. The advantage with this set-up, compared to designs based on ferromagnetic leads is that the dipolar and quadrupolar fields considered in Ref. \cite{Misiorny2013} here becomes vanishingly small. Therefore the effective isotropic electron mediated spin-spin interactions dominates the properties and control of the molecular dimer. In this sense, our system would serve as a representation of paramagnetic spintronics, or \emph{paratronics}.

\section{Methods}
\subsection{Magnetic molecular dimer in a junction}
\label{sec-dimer}
\begin{figure}[t]
\begin{center}
\includegraphics[width=0.99\columnwidth]{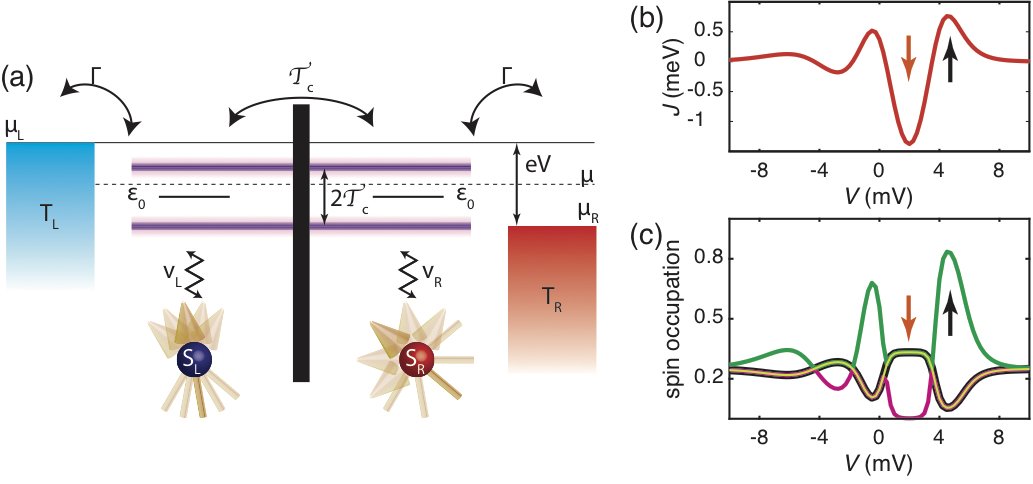}
\end{center}
\caption{(Color online) (a) Molecular dimer of paramagnetic molecules. An electron (at energy $\dote{0}$) in each molecule interacts with the localized spin moment ($\bfS_m$, $m=L,R$) via exchange ($v_m$) with the electron in the adjacent molecule (tunneling rate $\calT_c$) and with electron sin the left/right electrode (coupling $\Gamma$). The left/right nonmagnetic electrode is characterized by its electrochemical potential ($\mu_{L/R}$). Effective molecular orbutals ($\dote{0}\pm\calT_c$) emerge from intermolecular tunneling. (b) Effective exchange interaction between the localized spin moments as function of the voltage bias $V$. (c) Occupation of the states in the spin dimer. The green curve represents the occupation of the lowest energy eigenstate of the spin dimer which changes character between spin singlet and spin triplet states as function of the voltage bias. Other colors analogously represent the occupation of the consecutively higher energy eigenstates. In the region indicated by the red arrow three states are degenerate and form a spin triplet. Calculations are made at $T_L=4$ K and $T_R=9$ K for $\dote{0}-\mu=1$, $\calT_c=1$ meV, $v_m=5$ meV, and $\Gamma=1$, meV. In panels (b) and (c), the ferromagnetic and antiferromagnetic regimes of the spin dimer are indicated with red and black arrows, respectively.}
\label{fig-Fig1}
\end{figure}

The specific set-up we address in this article comprises a dimer of paramagnetic molecules which are embedded in series in the junction between normal metallic leads, see Figure \ref{fig-Fig1} (a). Each paramagnetic molecule comprises a localized spin moment which is embedded in a ligand structure, where the $sp$-orbitals define the spin-degenerate HOMO or LUMO orbitals. We assume that the $d$-, or, $f$-orbitals that constitute the molecular spins, hybridize only weakly with the $sp$-orbitals, allowing to consider the spin moment in the localized moment picture. As such, the localized moment interacts with the de-localized electrons only via local exchange.
We also neglect spin-orbit coupling in the molecular orbitals as well as considering them in single electron form, which is typically justified for the $sp$-electrons.
The molecular orbitals couple via tunneling both to one another and to the adjacent lead.

We model this set-up using the Hamiltonian
\begin{align}
\Hamil=&
	\Hamil_M
	+
	\Hamil_\text{int}
	+
	\Hamil_L
	+
	\Hamil_R
	+
	\Hamil_T.
\label{eq-model}
\end{align}
Here, the molecular HOMO or LUMO levels are defined by
\begin{align}
\Hamil_M=&
	\sum_\sigma
	\left(
		\sum_{m=L,R}\dote{m}\ddagger{m\sigma}\dc{m\sigma}
		+
		\calT_c(\ddagger{L\sigma}\dc{R\sigma}+H.c.)
	\right)
	,
\end{align}
where $\ddagger{m\sigma}$ ($\dc{m\sigma}$) creates (annihilates) an electron in the left ($L$) or right ($R$) molecule at the energy $\dote{m}=\dote{0}$ and spin $\sigma=\up,\down$, whereas $\calT_c$ defines the tunneling rate between the molecules. Internally in molecule $m$, the localized spin moment $\bfS_m$ interacts with the electron spin $\bfs_m=\sum_{\sigma\sigma'}\ddagger{m\sigma}\bfsigma_{\sigma\sigma'}\dc{m\sigma'}/2$, where $\bfsigma$ is the vector of Pauli matrices, via exchange
\begin{align}
\Hamil_\text{int}=&
	\sum_{m=L,R}v_m\bfs_m\cdot\bfS_m
	,
\end{align}
where $v_m$ is the exchange integral, and we assume that $v_m=v$.
We focus on the case with non-magnetic leads,
\begin{align}
\Hamil_{L/R}=&
	\sum_{\bfk\sigma\in L/R}\dote{\bfk}\cdagger{\bfk}\cc{\bfk}
	,
\end{align}
where $\cdagger{\bfk}$ creates an electron in the left ($L$; $\bfk=\bfp$) or right ($R$; $\bfk=\bfq$) lead at the energy $\dote{\bfk}$ and spin $\sigma$.
Tunneling between the leads and molecules is described by
\begin{align}
\Hamil_T=&
	\sum_{\bfp\sigma}\calT_L\cdagger{\bfp}\dc{L\sigma}
	+
	\sum_{\bfq\sigma}\calT_R\cdagger{\bfq}\dc{R\sigma}+H.c.
\end{align}
and we define the voltage bias $V$ across the junction by $eV=\mu_L-\mu_R$, where $\mu_\chi$, $\chi=L,R$, denotes the electrochemical potential of the lead $\chi$.

In this way $\Hamil_0=\Hamil_L+\Hamil_R+\Hamil_T+\Hamil_M$ provides a spin-degenerate background electronic structure which mediates the exchange interactions between the localized spin moments in $\Hamil_\text{int}$. The spin-degeneracy implies that these interactions are purely isotropic \cite{Fransson2014,Saygun2016a}, such that we retain the Heisenberg model only for the spins.

\subsubsection{Exchange interactions}
\label{ssec-exchange}
It has been shown that the Hamiltonian in Eq. (\ref{eq-model}) gives rise to the effective spin model \cite{Fransson2014}
\begin{align}
\Hamil_S=&
	\sum_{mn}
		\Bigl(
			J_{mn}\bfS_m\cdot\bfS_n
			+
			\bfD_{mn}\cdot[\bfS_m\times\bfS_n]
			+
			\bfS_m\cdot\mathbb{I}_{mn}\cdot\bfS_n
		\Bigr)
		.
\label{eq-HS}
\end{align}
Here, the parameter $J_{mn}$ denotes the isotropic Heisenberg interaction, whereas $\bfD_{mn}$ and $\mathbb{I}_{mn}$ respectively denotes the Dzyaloshinskii-Moriya and Ising interactions which both introduce anisotropic interactions into the system. In the present system, where the there is no spin-dependence imposed, either by external or internal forces, the anisotropic interactions vanish, $\bfD_{mn}=0$, $\mathbb{I}_{mn}=0$, and we retain the isotropic Heisenberg interaction only.

In this paper, we treat the spin dimer as a closed system in the sense that we require a conserved number of particles for which the occupations of the states are given by the Gibbs distribution. This is a valid approximation for localized spins where the hybridization with the surrounding itinerant electrons is small such that nature of the localized electrons can be described in terms of a Kondo-like model rather than a fluctuating spin model in the sense of the Hubbard or Anderson models.

The spin-spin interactions are, nevertheless, influenced by the tunneling current that flows through the molecular complex.
In such set-ups, the effective magnetic interaction parameter $J$ between the two spins can be calculated using the expression, see Refs. \cite{Fransson2014,Saygun2016a},
\begin{align}
J=&
	-\frac{\calT_c^2}{8\pi}
	v^2
	\sum_\chi
	\int
		\Gamma^\chi f_\chi(\omega)
		(\omega-\dote{0})
		\frac{(\omega-\dote{0})^2-\calT_c^2-(\Gamma/8)^2}{|(\omega-\dote{0}+i\Gamma/8)^2-\calT_c^2|^4}
	d\omega
	,
\label{eq-Jex}
\end{align}
where $\Gamma=\sum_{\chi=L,R}\Gamma^\chi$, with $\Gamma^\chi=2\pi\sum_{\bfk\sigma\in\chi}\calT_\chi^2\delta(\omega-\dote{\bfk})$,
is the coupling to the leads, and $f_\chi(\omega)$ is the Fermi function at the chemical potential $\mu_\chi$. Here, we have assumed that the molecular level $\dote{m}=\dote{0}$, $m=L,R$, and that the local exchange coupling $v_m=v$, which is reasonable for equivalent molecules.

We remark here that for a more general molecular assembly, that is, non-equivalent molecules and asymmetric couplings to the leads, the above expression for the exchange interaction should be replaced by the general formulas given in Ref. \cite{Fransson2014}. While this article is focused on dimers with equivalent molecules coupled symmetrically to the leads, we briefly discuss deviations from this case below.

\begin{figure}[t]
\begin{center}
\includegraphics[width=0.99\columnwidth]{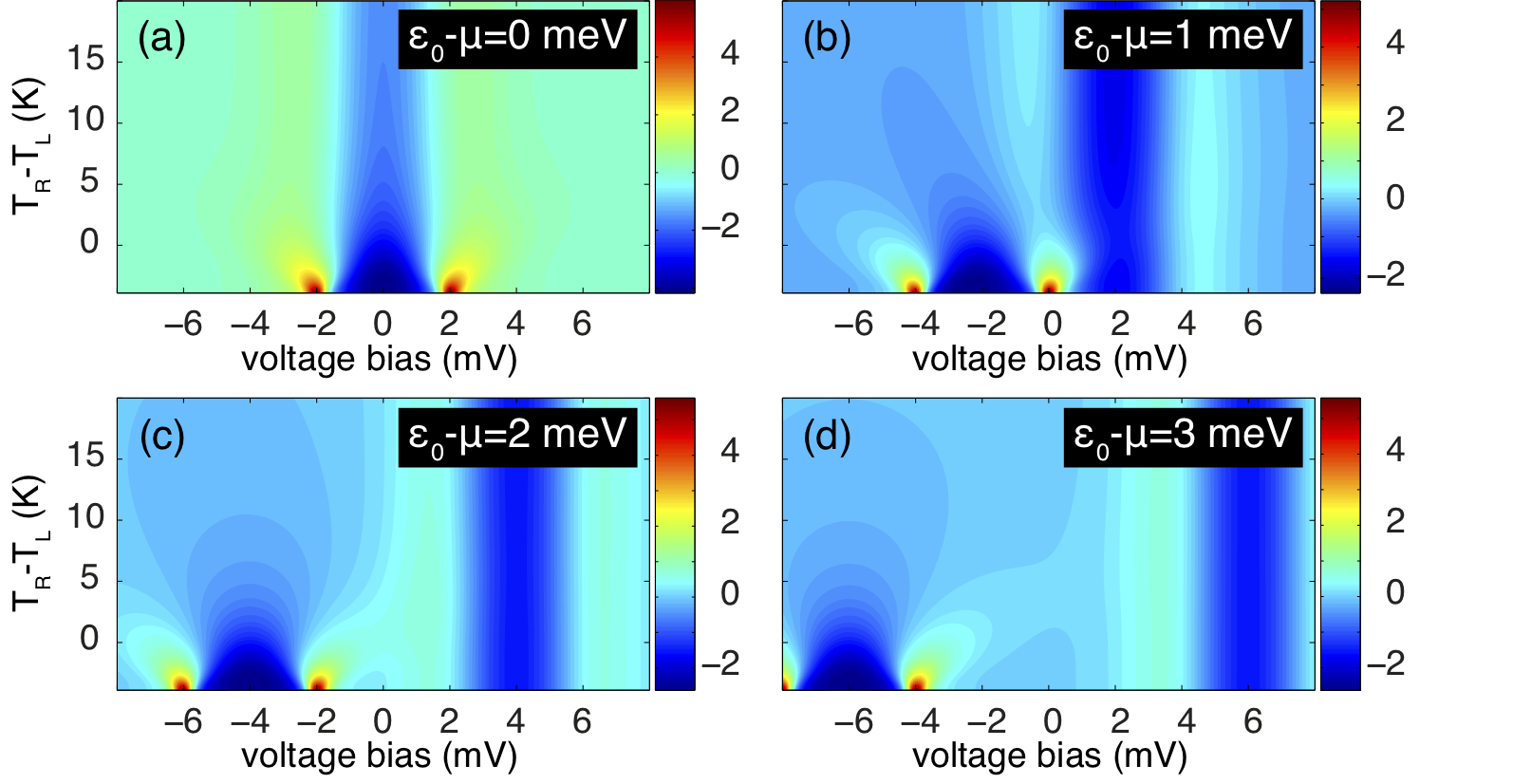}
\end{center}
\caption{(Color online) The exchange interaction $J$ (units: meV) between the molecular spins as function of voltage bias $V$ and temperature difference $\Delta T=T_R-T_L$ for different gating conditions, such that $\dote{0}-\mu=0,1,2,3$ meV. Other parameters are as in Figure \ref{fig-Fig1}.}
\label{fig-exchange}
\end{figure}

Notice that while the exchange interaction is defined in terms of the two-electron propagator $\eqgr{\bfs(t)}{\bfs(t')}$ \cite{Fransson2014}, we employ the de-coupling approximation $(-i){\rm sp}\bfsigma\bfG_{12}(t,t')\bfsigma\bfG_{21}(t',t)$. Here, $\bfG_{mn}(t,t')$ is a single-electron Green function for the molecular orbitals projected onto the sites $m$ and $n$, whereas ${\rm sp}$ defines the trace over spin $1/2$ space. This approximation is justified when neglecting local correlations; here the couplings to the localized spins. An obvious improvement would be to include these correlations and solve the two-electron propagator in a self-consistent scheme along with the Dyson equation for the single electron Green functions. However, since such an approach unavoidably also would include self-consistent calculations of the exchange and the spin configurations, and as we are mainly interested in qualitative effects, this is far beyond the scope of the present article.

\subsubsection{Non-equilibrium variations}
\label{ssec-ne}
The salient features of the voltage bias dependence of $J$ and the corresponding occupation of the spin states in the dimer are plotted in Figure \ref{fig-Fig1} (b), (c) for the case with $\dote{0}-\mu=1$ and a temperature difference of 5 K (for other parameters, see the figure caption). For the details about the interdependence between the electronic structure in the molecular orbitals and the spin-spin exchange we refer to Refs. \cite{Fransson2014,Saygun2016a}.
The temperatures of the leads are included in the respective Fermi function on which the exchange interaction within the spin dimer depends, see Eq. (\ref{eq-Jex}). In addition to the dependencies on the temperature of the leads, the exchange also depends on the energy of the molecular levels. These dependencies are illustrated in Figure \ref{fig-exchange}, which shows the exchange interaction energy as function of the voltage bias and temperature difference for different energies $\dote{0}-\mu$. There is a clear distinction between the situation in panel (a) compared to panels (b) -- (d), which originates in the level position relative to $\mu$.

\begin{figure}[t]
\begin{center}
\includegraphics[width=0.99\columnwidth]{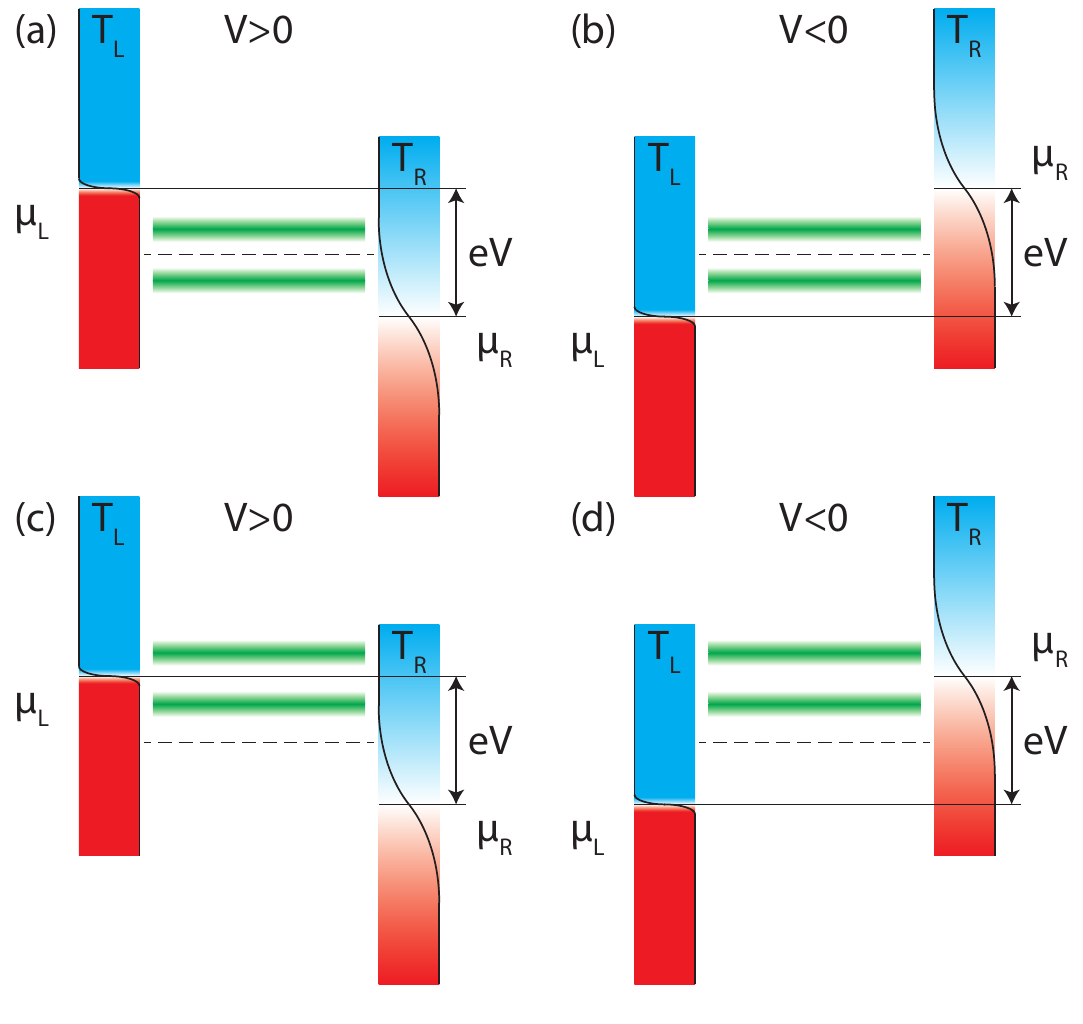}
\end{center}
\caption{(Color online) The combined role of the voltage bias and temperature difference on the tunneling conditions for (a), (c) positive and (b), (d) negative bias, and (a), (b) symmetric and (c), (d) asymmetric molecular orbitals around $\mu$ (dashed). The left (right) lead is defined at the temperature $T_L$ ($T_R$) and chemical potential $\mu_L$ ($\mu_R$); in the figure $T_L<T_R$ and $\mu_L-\mu_R=eV$. A low (high) temperature sustains a sharp (fuzzy) boundary between the occupied and unoccupied electron states in the metal. }
\label{fig-tempsketch}
\end{figure}

First, whenever the localized level $\dote{0}-\mu=0$, the molecular bonding (at the energy: $\dote{0}-\calT_c$) and anti-bonding (at the energy: $\dote{0}+\calT_c$) orbitals are centered symmetrically around $\mu$ which leads to that the influences from the left and right leads is symmetric with respect to voltage bias and independent of whether the source or drain electrode is warmer than the other. The situation is schematically depicted in Figure \ref{fig-tempsketch} (a), (b). The first thing to notice is a weakening of the exchange interaction caused by the increased thermal spread of the electrons in the hot reservoir, compared to the cold reservoir. In equilibrium, then, in comparison to the cold reservoir the hot reservoir contains a larger number of electrons above and below both molecular orbitals which contribute to an anti-ferromagnetic interaction and simultaneously a smaller number of electrons that contribute to the ferromagnetic exchange. This results in a weaker ferromagnetic interaction, which is verified in the simulations, see Figure \ref{fig-exchange} (a). The analogous argument holds for finite biases in the anti-ferromagnetic regime, however, utilized in the opposite way. Due to the thermal distribution of the electrons, there are electrons in the hot reservoir contributing to both the ferromagnetic and anti-ferromagnetic exchange which results in an effectively weaker anti-ferromagnetic exchange, see Figure \ref{fig-exchange} (a). Moreover, since the thermal distribution of the electrons is symmetric around the pertinent chemical potential, the effect is equal regardless of the polarity of the voltage bias across the junction, which basically indicates voltage bias symmetric transport properties of the junction under these conditions.

Next, whenever the molecule is gated such that the localized level $\dote{0}$ is off-set from the chemical potential $\mu$, a finite temperature difference do generate an asymmetry with respect to the voltage bias, see Figure \ref{fig-exchange} (b) -- (d). One can convince oneself that the exchange is symmetric with respect to the voltage bias at vanishing temperature difference, however, in general the asymmetry is conspicuous. Under the given conditions, $\dote{0}-\mu>0$ and variable $T_R$ with $T_L=4$ K fixed, the ferromagnetic regime at positive voltages remains nearly unaffected by changes in the temperature difference. This can be schematically understood from the sketch in Figure \ref{fig-tempsketch} (c), showing $\mu_L$ lying between the molecular orbitals. Then, the narrow thermal spread of the electrons in the left lead maintains a strong contribution to the ferromagnetic exchange while the contribution from the right lead is negligible more or less independently of its temperature. By the same token one can understand that the ferromagnetic regime for negative voltages is extremely sensitive to the temperature of the right lead, see Figure \ref{fig-tempsketch} (d). Under these conditions, the exchange interaction is dominated by the contribution from the right lead while the contribution from the left lead is negligible. Hence, at elevated temperatures there are competing contributions to the exchange of both ferromagnetic and anti-ferromagnetic nature which leads to a very weak ferromagnetic exchange that weakens with increasing temperature difference.

\subsubsection{Deviations from perfectly equivalent and symmetric dimers}
\label{ssec-asymmetric}
In realistic situations one can expect that the local parameters vary from molecule to molecule, although they are meant to be considered as equivalent. Here, we briefly discuss some implications on the indirect exchange interaction under different local exchange integrals $v_m$, finite level off-set $\Delta=\dote{L}-\dote{R}$, between the molecular levels $\dote{m}$, and asymmetric couplings to the leads.

For instance, for non-equivalent molecules such that $\dote{1}\neq\dote{2}$, the exchange interaction $J$ tends to become strongly asymmetric with respect to the voltage bias, see for instance Ref. \cite{Saygun2016a}. It may even lead to situations where $J$ is ferromagnetic (negative) for one polarity of the voltage and anti-ferromagnetic (positive) for the opposite. In those situations, one would expect the resulting transport properties to be significantly different for the two voltage polarities, something that was also verified by a strong rectification property \cite{Saygun2016a}.

Except for the strong dependence of the exchange interaction on the level off-set between the two molecules, the dimer structure is remarkably robust against small variations and asymmetries in the couplings $\Gamma^\chi$ to the leads as well as to inequalities in the local Kondo exchange couplings $v_m$. Differences of up to $20\ \%$ between the couplings of the two molecules do not leads to any essential variations in the exchange interaction and are, therefore, not expected to be detrimental to the transport properties either.

\subsubsection{Spin expectation values}
\label{sse-spinexp}
For later purpose we introduce the notation $\av{S_\chi^z}$ for the expectation values which are projections onto the molecule $\chi$ of the total spin expectation value $\av{S}$ of the dimer. The expectation value $\av{S}$ is calculated with respect to the spin Hamiltonian $\Hamil_S=J\bfS_L\cdot\bfS_R$, with $J$ obtained from Eq. (\ref{eq-Jex}). As we are considering the individual moments to remain in their ground states under all conditions, the restricted Hilbert space corresponding to this set-up is four dimensional labeled by the states, say, $\{\ket{\up\up},\ket{\up\down},\ket{\down\up},\ket{\down\down}\}$, where the first (second) entry corresponds to the left (right) spin. Then, the projected expectation value is calculated by the expression $\av{S_\chi^z}\equiv\bra{\sigma\sigma'}\exp\{-\beta_\text{av} E_{\sigma\sigma'}\}S_\chi^z\ket{\sigma\sigma'}/\bra{\sigma\sigma'}\exp\{-\beta_\text{av} E_{\sigma\sigma'}\}\ket{\sigma\sigma'}$, where the operator $S_\chi^z$ is expressed in the basis of the dimer. Here, also $1/\beta_\text{av}=k_B(T_L+T_R)/2$, represents the average temperature of the two leads. Although this appears to be a severe simplification, it turns out that the spin occupation numbers vary slowly as function of the temperature in the current set-up, which justifies this simplistic treatment.

We remark here that any temperature mediated by the tunneling electrons vanish under the present condition in the employed approximation. To clarify this point, we notice that the effective spin model introduced in Eq. (\ref{eq-HS}) in principle also contains the contribution
\begin{align}
\Hamil_\tau=&
	-g\mu_B\bfB_\tau\cdot\sum_m\bfS_m
	,
\end{align}
where the magnetic field $\bfB_\tau$ is proportional to the current through the junction, see, e.g., Refs. \cite{Fransson2008,Chudnovskiy2008,Hammar2016,Fransson2014,Ludwig2017,arXiv:1702.02820}. This field provides a torque on the loclized spin, however, only when the spin-degeneracy is broken in the current. In other words, this current induced magnetic field vanishes whenever the tunneling current is spin-degenerate. Hence, since our set-up is spin-degenerate by construction, the current induced magnetic field does not generate any renormalization of the localized spin and, in particular, the temperatures of the leads are not transmitted via the charge current to the spins.

\subsection{Transport formulas}
\label{sec-currents}
Here, we will briefly go through and summarize the approach we employ to study the transport properties governed by the molecular spin dimer. 
First, the currents under interest are the charge and thermal currents, which are defined for the flows through interface $\chi$ as
\begin{subequations}
\begin{align}
I_c^\chi=&
	-eI_N^\chi
	,
\\
I_Q^\chi=&
	I_E^\chi-\mu I_N^\chi
	,
\end{align}
\end{subequations}
respectively, where $I_{E(N)}^\chi$ denotes the energy (particle) current through the junction.
Second, we notice that the energy and particle currents through the system are defined as the rate of change of the energy and the particle number, respectively.
Concerning the particle flux, conservation of particles ensures that the rate of particles passing through one of the interfaces equals the corresponding rate at the other interface. However, the component in the thermal current that is generated by the particle flux is not a conserved quantity \cite{Galperin2007}. This problem can be avoided by considering the local entropy production rate in the molecular dimer, expressed through $I_Q=I_Q^L-I_Q^R=(I_E^L-\mu_LI_N^L)/2-(I_E^R-\mu_RI_N^R)/2$. Then, since the particle contribution can be written as $-(\mu_L+\mu_R)I_N^L/2=-\mu I_N^L/2$ and since we measure all energies relative to the chemical potential $\mu$, which is effectively zero, the particle contribution to the thermal current vanishes.

We remark here that the terminology \emph{entropy production rate} for the quantity $I_Q$ is justified in the stationary regime since then the heat current $I_Q^{\chi}$, $\chi=L,R$, flowing through the left or right interface equals the corresponding entropy flow $I_S^\chi$ multiplied by the temperature, $T_\chi I_S^\chi$. This can be understood by considering that the rate of change of the non-equilibrium grand potential vanishes in the stationary regime, $\dt\Omega=0$, which yields $T\dt S=\dt\av{\Hamil}-\mu\dt\av{N}$. On the one hand, the entropy production rate is given by the difference of the entropies flowing through the left and right interface, $\dt S^L-\dt S^R$, with dimensions energy divided by temperature and time. On the other hand, the well defined and calculable quantity is here defined by the thermal currents $I_Q^\chi$ and, due to the strong intimacy between the two quantities $\dt S^\chi$ and $I_Q^\chi$, we shall use the term energy production rate for $I_Q$ with dimensions energy divided by time.

We express the fluxes at the interface $\chi$ in the generic forms
\begin{subequations}
\begin{align}
I_E^\chi=&
	\frac{d}{dt}\av{\Hamil_\chi}
	=
	-2\calT_\chi\im\sum_{\bfk\sigma}\dote{\bfk}(-i)F_{\bfk\chi\sigma}^<(t,t)
	,
\\
I_N^\chi=&
	\frac{d}{dt}\av{N_\chi}
	=
	-2\calT_\chi\im\sum_{\bfk\sigma}(-i)F_{\bfk\chi\sigma}^<(t,t)
	,
\end{align}
\end{subequations}
where we have defined the lesser form $F^<_{\bfk\chi\sigma}(t,t')$ of the transfer Green function $F_{\bfk\chi\sigma}(t,t')=\eqgr{\cc{\bfk}(t)}{\ddagger{\chi\sigma}(t')}$. Using standard theory \cite{Jauho1994} and restoring $\hbar$, the currents can be compactly written on the form
\begin{subequations}
\label{eq-IEN}
\begin{align}
I_E^\chi=&
	(-i)
	\frac{\Gamma^\chi}{\hbar}
	{\rm sp}
	\int
		\omega
			\Bigl(
				f_\chi(\omega)\bfG^>_\chi(\omega)
				+
				f_\chi(-\omega)\bfG^<_\chi(\omega)
			\Bigr)
    	\frac{d\omega}{2\pi}
	,
\label{eq-IE}
\\
I_N^\chi=&
	(-i)
	\frac{\Gamma^\chi}{\hbar}
	{\rm sp}
	\int
			\Bigl(
				f_\chi(\omega)\bfG^>_\chi(\omega)
				+
				f_\chi(-\omega)\bfG^<_\chi(\omega)
			\Bigr)
	\frac{d\omega}{2\pi}
	.
\label{eq-IN}
\end{align}
\end{subequations}
Here, ${\rm sp}$ denotes the trace over spin 1/2 space whereas $\bfG^{</>}_\chi$, $\chi=L,R$, denotes the $2\times2$-matrix Green function of the molecule adjacent to the lead $\chi$.

\subsubsection{Derivation of the molecular Green function}
\label{sec-GF}
We make further analytical progress by constructing an explicit expression for the Green function $\bfG_\chi$. To this end we include the broadening effects from the couplings to the leads as well as the inter-molecular tunneling and intra-molecular exchange interactions with the localized spin moments. These presumptions lead to that we can write the retarded/advanced Green functions $\bfG_\chi$ weighted on molecule $\chi$ as \cite{Saygun2016a}
\begin{align}
\bfG_{L(R)}^{r/a}(\omega)=&
	\frac{1}{2\tilde\calT_c}
	\sum_{s=\pm}
		\frac{\tilde\calT_c\sigma^0-2sv\av{S_{R(L)}^z}\sigma^z}
		{\omega-E_s\pm i\Gamma/8}
	,
\end{align}
where the excitation energies $E_\pm=\dote{0}\pm\tilde\calT_c/2$ and $\tilde\calT_c^2=v^2\av{S_L^z-S_R^z}^2+4\calT_c^2$.

The Green function $\mathbb{G}$ for the full dimer system is a $4\times4$-matrix $\mathbb{G}=\{\bfG_{\chi\chi'}\}_{\chi,\chi'=L,R}$, in which each entry is a $2\times2$-matrix $\bfG_{\chi\chi'}=\{G_{\chi\sigma \chi\sigma'}\}_{\sigma,\sigma'=\up,\down}$. Here, the subscripts $\chi,\chi'$ refer to the left (right) molecule if $\chi\chi'=LL$ ($\chi\chi'=RR$) and coupling between the molecules for $\chi\chi'=LR$ or $\chi\chi'=RL$. For brevity, we write $\bfG_{\chi\chi}=\bfG_\chi$. Each entry is defined by
\begin{align}
G_{\chi\sigma \chi'\sigma'}(z)=&
	\av{\inner{\dc{\chi\sigma}}{\ddagger{\chi'\sigma'}}}(z)
\nonumber\\=&
	\int
		\eqgr{\dc{\chi\sigma}(t)}{\ddagger{\chi'\sigma'}(t')}
		e^{-iz(t-t')}
	dt'
	.
\end{align}
The equation of motion for $\mathbb{G}$ can be summarized in the Dyson equation
\begin{align}
\mathbb{G}=&
	\mathbb{G}_0
	+
	\mathbb{G}_0
	\bfSigma
	\mathbb{G}
	,
\end{align}
where $\mathbb{G}_0$ is the bare Green function for the coupled molecules, however, without couplings to the leads, whereas $\bfSigma$ defines the self-energy generated by the couplings to the leads. It can be noticed that since molecule 1 (2) only couples to the left (right) lead, the retarded form of this self-energy can be written as the diagonal matrix $\bfSigma=(-i)\diag{\Gamma^L_\up,\ \Gamma^L_\down,\ \Gamma^R_\up,\ \Gamma^R_\down}{}/2$. Considering spin-degenerate conditions, we can set $\Gamma_\sigma^\chi=\Gamma^\chi/2$. As an effect of the Dyson equation for $\mathbb{G}$, the corresponding lesser/greater forms are given by $\mathbb{G}^{</>}=\mathbb{G}^r\bfSigma^{</>}\mathbb{G}^a$, where the lesser/greater forms of the self-energy is given by
\begin{align}
\bfSigma^{</>}(\omega)=&
	(\pm i)
	\frac{1}{4}
	\begin{pmatrix}
		f_L(\pm\omega)
			\Gamma^L
			\sigma^0
		& 0 \\
		0 &
		f_R(\pm\omega)
			\Gamma^R
			\sigma^0
	\end{pmatrix}
	.
\end{align}

\begin{figure}[t]
\begin{center}
\includegraphics[width=\columnwidth]{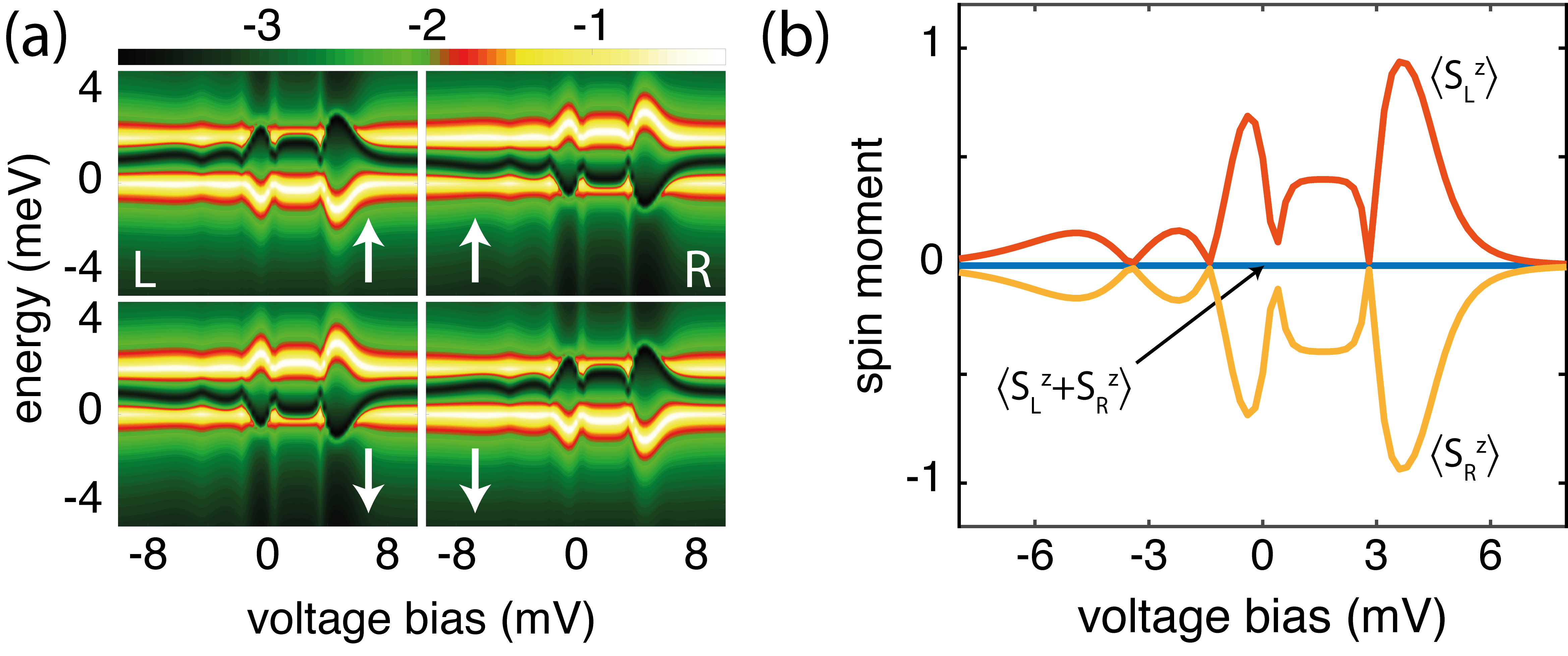}
\end{center}
\caption{(Color online) (a) Molecule $(L$, $R$) and spin projected (indicated by white arrows) densities of electron states of the left and right non-gated ($\mu=0$) molecules, calculated using $G_{LL(RR)\sigma}$ given in Eq. (\ref{eq-fullGreenFunction}), respectively, as function of the voltage bias $V$ and energy $\omega$. (b) Molecule projected {eq-fullGreenFunction} of the localized spins $\av{S_\chi^z}$ and the total magnetic moment $\av{S_L^z+S_R^z}$. Parameters are as in Figure \ref{fig-Fig1}.}
\label{fig-DOS}
\end{figure}

In order to find the analytical forms of the local Green function, we notice that since the spin degrees of freedom are uncoupled, we can write $\mathbb{G}$ in block diagonal form. In this representation, the blocks are distinguished by the spin index whereas each block can be written on the form
\begin{widetext}
\begin{align}
\Bigl\{G_{\chi\chi'\sigma}^{r/a}(\omega)\Bigr\}_{\chi,\chi'=L,R}
\nonumber\\=&
	\left[
		\begin{pmatrix}
			\omega-\dote{0}\pm i\Gamma^L/4 & -\calT_c \\
			-\calT_c & \omega-\dote{0}\pm i\Gamma^R/4
		\end{pmatrix}
	\right.
	\left.
		-
		v\sigma^z_{\sigma\sigma}
		\begin{pmatrix}
			\av{S_L^z} & 0 \\
			0 & \av{S_R^z}
		\end{pmatrix}
	\right]^{-1}
=
	\frac{	\begin{pmatrix}
		\omega-\dote{0}-v\av{S_R^z}\pm i\Gamma^R/4 & \calT_c \\
		\calT_c & \omega-\dote{0}-v\av{S_L^z}\pm i\Gamma^L/4
	\end{pmatrix}
	}{(\omega-E_+^{r/a})(\omega-E_-^{r/a})}
	,
\label{eq-fullGreenFunction}
\end{align}
\end{widetext}
where $E_s^{r/a}=\dote{0}+s\tilde\calT_c/2\mp i(\Gamma^L+\Gamma^R)/8=\dote{0}+s\tilde\calT_c/2\mp i\Gamma/8$ and where we have put $\Gamma^\chi=\Gamma/2$ in the last equality. Accordingly, the Green function weighted on the left molecule with spin $\sigma$ is, therefore, given by the entry $G_{LL\sigma}^{r/a}(\omega)$, which can also be written on the form
\begin{align}
G_{L\sigma}^{r/a}(\omega)=&
	\frac{1}{2\tilde\calT_c}
	\sum_{s=\pm}
		\frac{\tilde\calT_c-2sv\av{S_R^z}\sigma^z_{\sigma\sigma}}{\omega-E_s\pm i\Gamma/8}
	,
\end{align}
with $E_\pm=\dote{0}\pm\tilde\calT_c/2$.

When putting these results into the combination of the lesser and greater Green functions in the transport formulas, Eq. (\ref{eq-IEN}), we retain only
\begin{align}
f_\chi(\omega)\bfG^>_\chi(\omega)-&
	f_\chi(-\omega)\bfG^<_\chi(\omega)
\nonumber\\&\hspace{-1cm}=
(-i)\frac{\Gamma^R}{2}
	\Bigl(
		f_\chi(\omega)
		-
		f_{\chi'}(\omega)
	\Bigr)
		\bfG^r_{\chi\chi'}(\omega)\bfG^a_{\chi'\chi}(\omega)
	,
\end{align}
out of which the only finite terms are the ones that couple the two molecules to one another and it is, therefore, necessary to study the forms of the site off-diagonal Green functions $\bfG_{LR}$ and $\bfG_{RL}$.
By a straight forward calculation one finds that these Green functions can be written as
\begin{align}
\bfG_{\chi\chi'}^{r/a}(\omega)=&
	\frac{\calT_c}{2}
	\sum_{s=\pm1}
	\frac{\sigma^0-s\sigma^z}{\omega-\dote{0}-sv\av{S_\chi^z}\pm i\Gamma/8}
	\bfG_{\chi'}^{r/a}(\omega)
\nonumber\\=&
	\frac{\calT_c}{4\tilde\calT_c}
	\sum_{ss'=\pm1}
		\frac{[\sigma^0-s\sigma^z][\tilde\calT_c\sigma^0-2s'v\av{S_\chi^z}\sigma^z]}
		{[\omega-\dote{0}-sv\av{S_\chi^z}\pm i\Gamma/8][\omega-E_{s'}\pm i\Gamma/8]}
	.
\end{align}

\section{Results and discussion}
\label{sec-results}

\begin{figure}[t]
\begin{center}
\includegraphics[width=0.99\columnwidth]{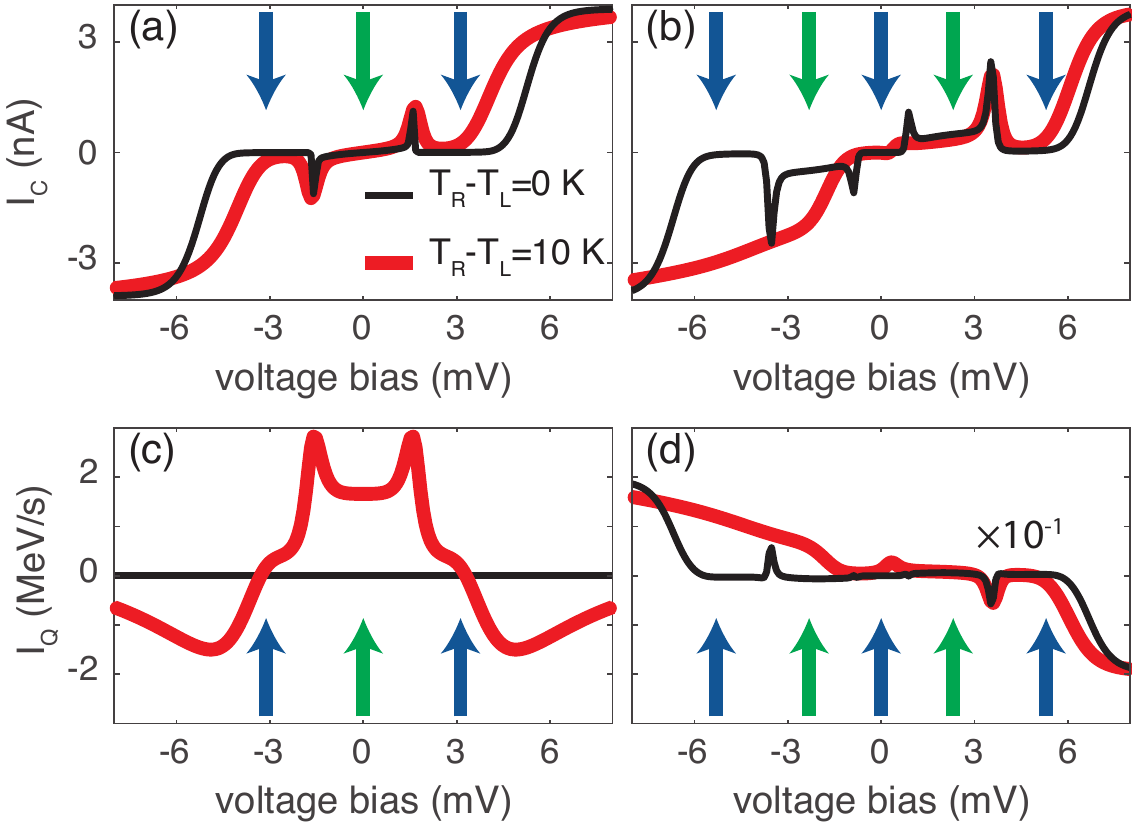}
\end{center}
\caption{(Color online) (a), (b), The charge current $I_C$ (units: nA) and (c), (d), entropy production rate $I_Q$ (units: MeV/s) as function of voltage bias $V$ for $T_R-T_L=0$ (black -- faint) and $T_R-T_L=10$ K (red -- bold) for (a), (c), $\dote{0}-\mu=0$ meV and (b), (d),  $\dote{0}-\mu=1$ meV. Other parameters are as in Figure \ref{fig-Fig1}. In panels (a) and (b), the ferromagnetic and anti-ferromagnetic regimes are indicated with green and blue arrows, respectively.}
\label{fig-Ictraces}
\end{figure}

Regarding the charge transport across the junction, most of its properties can be understood in terms of the degree of delocalization of the electronic density in the molecular dimer. As have been discussed in a previous publication \cite{Saygun2016a}, the magnetic states and configurations lead to qualitatively distinct regimes of the electronic properties of the dimer, something that is conveyed over to the inherit transport properties of the molecular dimer. Accordingly, the conductance in the ferromagnetic regime is expected to better, or, higher than in the anti-ferromagnetic regime. This conjecture is based on the fact that in the anti-ferromagnetic regime the spin projected electronic density in the bonding state is strongly localized to one \emph{or} the other molecule, see Figure \ref{fig-DOS} (a), and in the opposite fashion for the anti-bonding orbital. Therefore, an electron residing in one of the molecules has a nearly vanishing probability to tunnel over the other molecule which leads to a strongly suppressed conductance. In the ferromagnetic regime, however, the electronic density is more delocalized throughout the molecular dimer which allows for resonant tunneling between the molecules and, in turn, leads to an increased conductance. These features and disparities of the two magnetic regimes can be observed in the current, see Figure \ref{fig-Ictraces} (a), (b), showing the charge current $I_c$ as function of the voltage bias at zero temperature difference (black -- faint) and $T_R-T_L=10$ K (red -- bold). Especially for vanishing temperature difference, the anti-ferromagnetic and ferromagnetic regimes (indicated by blue and green arrows, respectively) are strikingly separated by sharp current resonances which originates from a vanishing exchange between the spin. The absence of the exchange interaction leads to uncorrelated spins and a completely delocalized charge density in the molecular dimer, which allows for a large current flow in the same way as in the high voltage regime where $J\rightarrow0$.

\begin{figure}[t]
\begin{center}
\includegraphics[width=0.99\columnwidth]{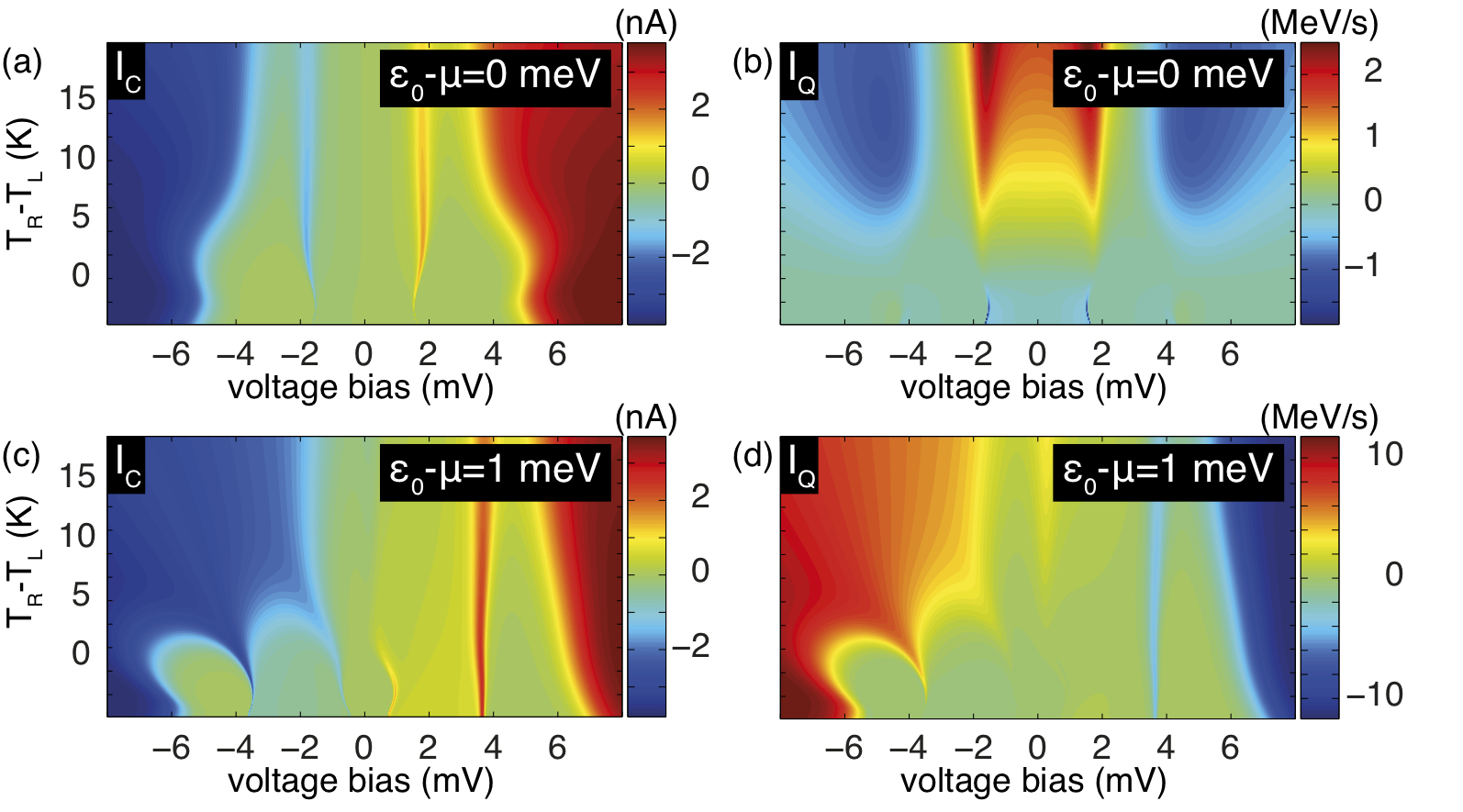}
\end{center}
\caption{(Color online) (a), (c), The charge current $I_C$ (units: nA) and (b), (d), entropy conductance $I_Q$ (units: MeV/s) as function of voltage bias $V$ and temperature difference $T_R-T_L$ for (a), (b), $\dote{0}-\mu=0$ meV and (c), (d),  $\dote{0}-\mu=1$ meV. Other parameters are as in Figure \ref{fig-Fig1}}
\label{fig-currents}
\end{figure}

At zero temperature difference across the junction, the current-voltage characteristics is necessarily symmetric whenever then electronic structure of the molecular dimer are symmetrically distributed, as is the case we consider here. Introduction of a finite temperature difference changes this scenario, however, a necessary condition for breaking the symmetric current-voltage characteristics is that the bonding and anti-bonding orbitals do not surround $\mu$ symmetrically. This can be seen in the red -- bold traces in Figure \ref{fig-Ictraces}. In panel (a), the molecular level $\dote{0}=\mu$ which accordingly leads to a symmetric current as function of the voltage bias. In panel (b), in contrast, the molecular level satisfies $\dote{0}-\mu=1$ meV, which then generates a striking asymmetry in the current for finite temperature differences. This result can be traced back to the dramatically changed, weakened, exchange interaction between the spins which causes an increased degree of delocalization of the electronic density in the molecular dimer. In effect, therefore, both the ferromagnetic and anti-ferromagnetic regimes are essentially destroyed for negative voltages, see Figure \ref{fig-Ictraces} (b) (red -- bold). Hence, under the temperature difference between the leads the molecular dimer functions partially as a rectifying device where the magnetically active regimes can be employed in the forward direction while the system behaves like a normal conductor in the backward.
In Figure \ref{fig-currents} (a), (c), we show contour plots of the charge current as function of the voltage bias and temperature difference for (a) $\dote{0}-\mu=0$ meV and (c) $\dote{0}-\mu=1$ meV. The properties discussed in detail above can be seen to be verified in the larger picture, varying continuously with the variations of the external conditions. Especially in the latter case ($\dote{0}-\mu=1$ meV), the magnetically active regimes for negative voltage biases are seen to be destroyed already for small temperature differences, closely following the behavior of the exchange interactions, c.f., Figure \ref{fig-exchange}.

Next, in the discussion of the entropy production rate, we again notice that most of the expected features can be explained in terms of the properties of the electronic structure in the molecular dimer analogous as with the charge current. However, for vanishing temperature difference between the leads a clear distinction compared to the charge current is that the finiteness of the entropy production rate strongly depends on the energy of the localized level in the molecular assembly. This can be traced back to the product of the energy and the distribution functions ($\bfG^{</>}_\chi$) of the molecular dimer in, for instance, Eq. (\ref{eq-IE}). One can make a simple comparison between the qualitative properties of the two currents by assuming a Lorentzian model, $1/[(\omega+\mu-\dote{0})^2+(\Gamma/2)^2]$, for the device embedded between the leads at low temperatures. Then, under the voltage bias $\mu_{L/R}=\mu\pm eV/2$, the two currents behave as
\begin{subequations}
\label{eq-IcIq}
\begin{align}
I_c\sim&
	\frac{1}{\Gamma/2}
	\biggl(
		\arctan\frac{eV/2-\dote{0}}{\Gamma/2}
		+
		\arctan\frac{eV/2+\dote{0}}{\Gamma/2}
	\biggr)
	,
\\
I_Q\sim&
	\frac{1}{2}
	\ln\frac{(eV/2-\dote{0})^2+(\Gamma/2)^2}{(eV/2+\dote{0})^2+(\Gamma/2)^2}
\nonumber\\&
	+
	\frac{\dote{0}-\mu}{\Gamma/2}
	\biggl(
		\arctan\frac{eV/2-\dote{0}}{\Gamma/2}
		+
		\arctan\frac{eV/2+\dote{0}}{\Gamma/2}
	\biggr)
	.
\end{align}
\end{subequations}
Hence, while both the charge current and entropy production rate have normal on-sets associated with the energy $\dote{0}$, described by the $\arctan{}$-component, the entropy production rate is also logarithmically resonant at $\dote{0}$. In addition, for the entropy production rate the on-set at $\dote{0}$ is weighted by the position of $\dote{0}$ relative to $\mu$ and, therefore, this contribution to the entropy production rate is expected to the small whenever $\dote{0}$ lies in the vicinity of $\mu$. The logarithmic contribution tends to be small for large voltage biases, since it leads to an increasingly symmetric argument, and is accordingly only significant for voltages such that either $eV/2-\dote{0}$ or $eV/2+\dote{0}$ is small.

Extrapolating this discussion to our present case with the molecular dimer, we can verify these expected features under the conditions of vanishing temperature difference. This can be be seen in Figure \ref{fig-Ictraces} (c), (d) (black -- faint). At $\dote{0}-\mu=0$, panel (a), the entropy production rate is identically zero while a finite off-set from $\mu$, panel (b), yields a finite entropy production rate. In the latter case, the entropy production rate is small in the magnetically active regime and grows large only at large voltages where the exchange interaction vanishes.
The entropy production rate tends to be efficiently transmitted only when the molecular dimer is in its fully conjugated state, that is, when the exchange interaction is small. This is contrast to the charge current where the less localized electronic density, in the molecular dimer, in the ferromagnetic regime yields a significant difference compared to the anti-ferromagnetic.

The reason for this qualitative difference between the charge and entropy production rate, see Eq. (\ref{eq-IcIq}), can be found in the properties of the two currents in the ferromagnetic regimes. In particular, the logarithmically resonant contribution to the entropy production rate, which is absent in the charge current, peaks near the energies of the molecular bonding and anti-bonding orbitals, something which occurs within the ferromagnetic but not in the anti-ferromagnetic regime. Thereto, these resonant peaks are oppositely directed compared to the contributions varying like $\arctan{x}$.
As for the entropy production rate under finite temperature difference between the leads, see Figure \ref{fig-Ictraces} (c), (d) (red -- bold) and Figure \ref{fig-currents} (b), (d), this behavior is verified. This feature can be traced back to an increased degree of delocalization in the ferromagnetic regime due to the increased thermal broadening from the hotter reservoir.

\begin{figure}[t]
\begin{center}
\includegraphics[width=0.99\columnwidth]{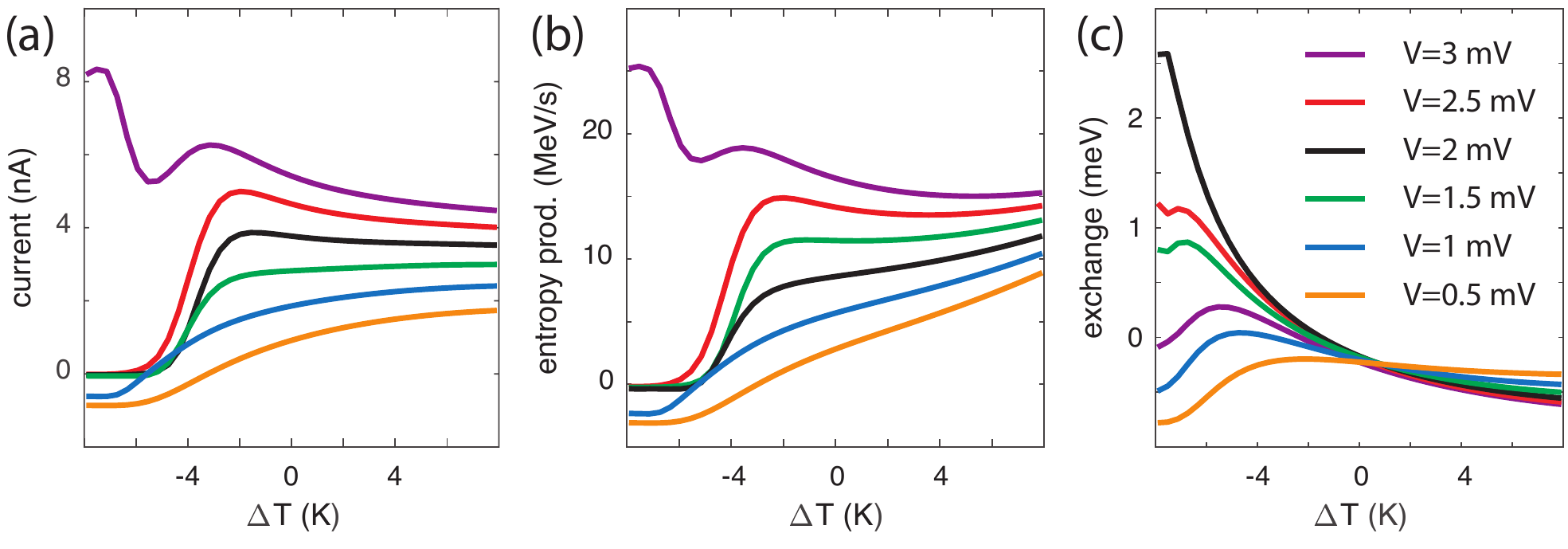}
\end{center}
\caption{(Color online)
Thermal rectification. (a) Charge current, (b) entropy production rate, and (c) exchange interaction, as function of the temperature difference for different voltage biases. Here, $\dote{0}-\mu=2$ and $\calT_c=3$ meV, while other parameters are as in Figure \ref{fig-Fig1}.
}
\label{fig-ThermRect}
\end{figure}

In the case where the localized level is off-set from $\mu$, there is an interesting observation to be made in the temperature and voltage dependence of both the charge and entropy production rate. At finite temperature difference and voltage bias, there is a strong asymmetry with respect to zero voltage which is clear signature that the magnetically active dimer potentially can be used for current rectification. Previously it was shown in Ref. \cite{Saygun2016a} that the charge current can be rectified by introducing a level off-set between the localized levels in the two molecules, something which possibly can be obtained by using different types of molecules. The asymmetry induced by the temperature difference provides a different mechanism to rectification which is also viable for the entropy production rate. Indeed, the plots in Figure \ref{fig-currents} (c), (d), illustrate that both the charge and the entropy production rate is rectified at finite temperature differences upon changing the polarity of the voltage bias. For this to be successful, the dimer has to be set-up with a finite level off-set from $\mu$ and a finite voltage bias. Moreover, the parameters of the molecular dimer have to be tuned such that the off-set between the bonding and anti-bonding orbitals is larger than the thermal energy fed into the dimer from the hotter electrode. This enables a crossover from the anti-ferromagnetic regime to either the ferromagnetic or paramagnetic regime, under changes in the temperature difference, which leads to strong variations in the currents. In principle, the stronger anti-ferromagnetic exchange one can obtain for one polarity of the temperature difference and the weaker the exchange can be in the opposite, regardless of sign, the larger the ratio between the currents for the two polarities. In Figure \ref{fig-ThermRect} we plot the (a) charge and (b) entropy production rate as function of the temperature difference for different voltage biases. We apply half the difference to the temperature in each lead such that $T_L=T-\Delta/2$ and $T_R=T+\Delta/2$, where the base temperature $T=4$ K. The calculations confirm the argument that variations of the exchange interaction from strongly anti-ferromagnetic to weakly ferromagnetic [$V\approx2$ mV, see Figure \ref{fig-exchange} (c)] indeed yields the larger ratio between the large and small current regimes.

\begin{figure}[t]
\begin{center}
\includegraphics[width=0.99\columnwidth]{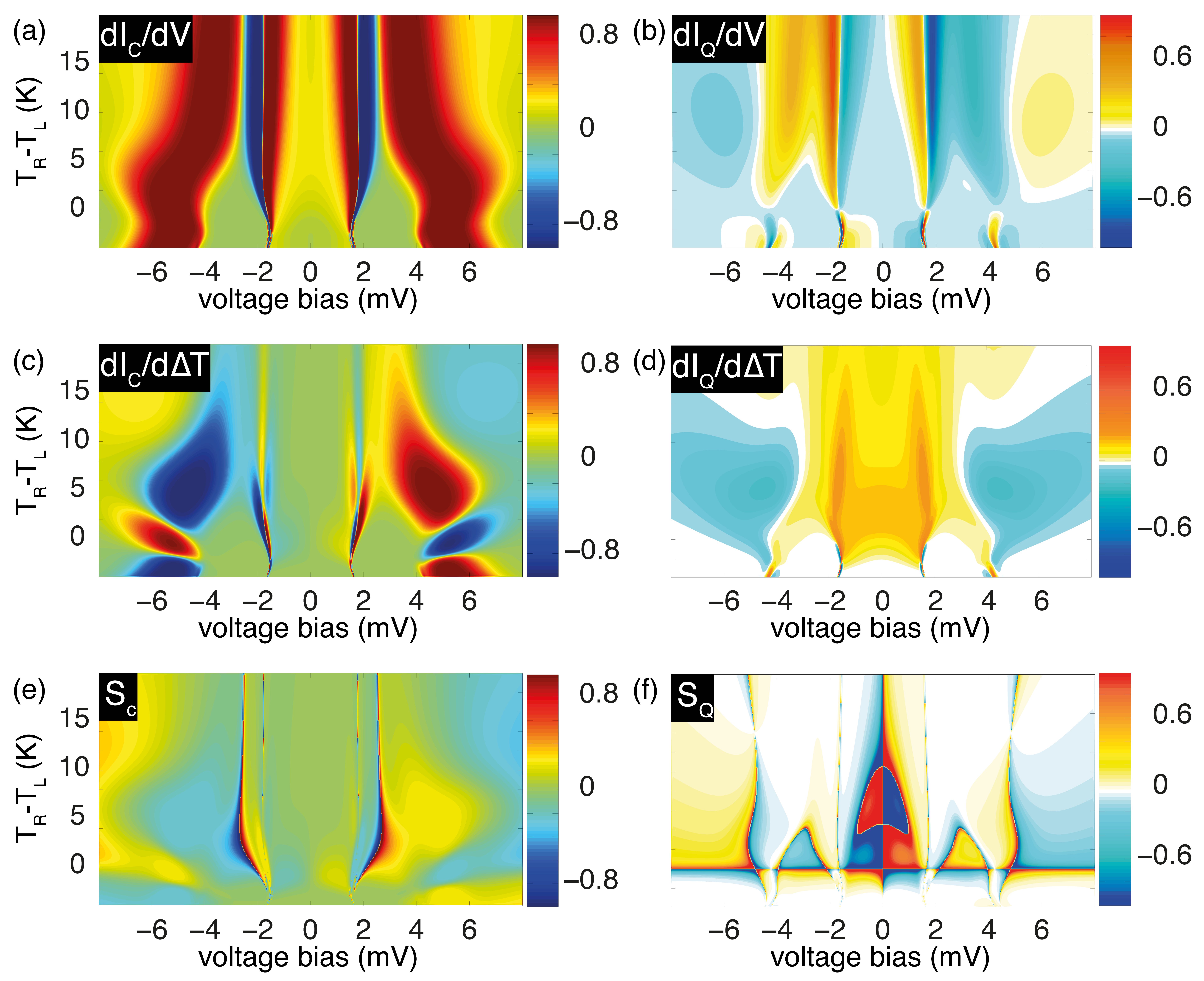}
\end{center}
\caption{(Color online)
(a) -- (d) Differential charge conductance (a), (c), and entropy production rate (b), (d), with respect to the voltage bias (a), (b), and temperature difference (c), (d).
(e) $S_c$ and (f) $S_Q$.
Here, $\dote{0}-\mu=0$, while other parameters are as in Figure \ref{fig-Fig1}.
All contours show $(1+\exp\{\sigma{\cal F}\})^{-1}$, where ${\cal F}$ is one of $\partial I_x/\partial V$, $\partial I_x/\partial\Delta T$, and $S_x$, whereas $\sigma$ is a scaling parameter.
}
\label{fig-conde0}
\end{figure}

\begin{figure}[t]
\begin{center}
\includegraphics[width=0.99\columnwidth]{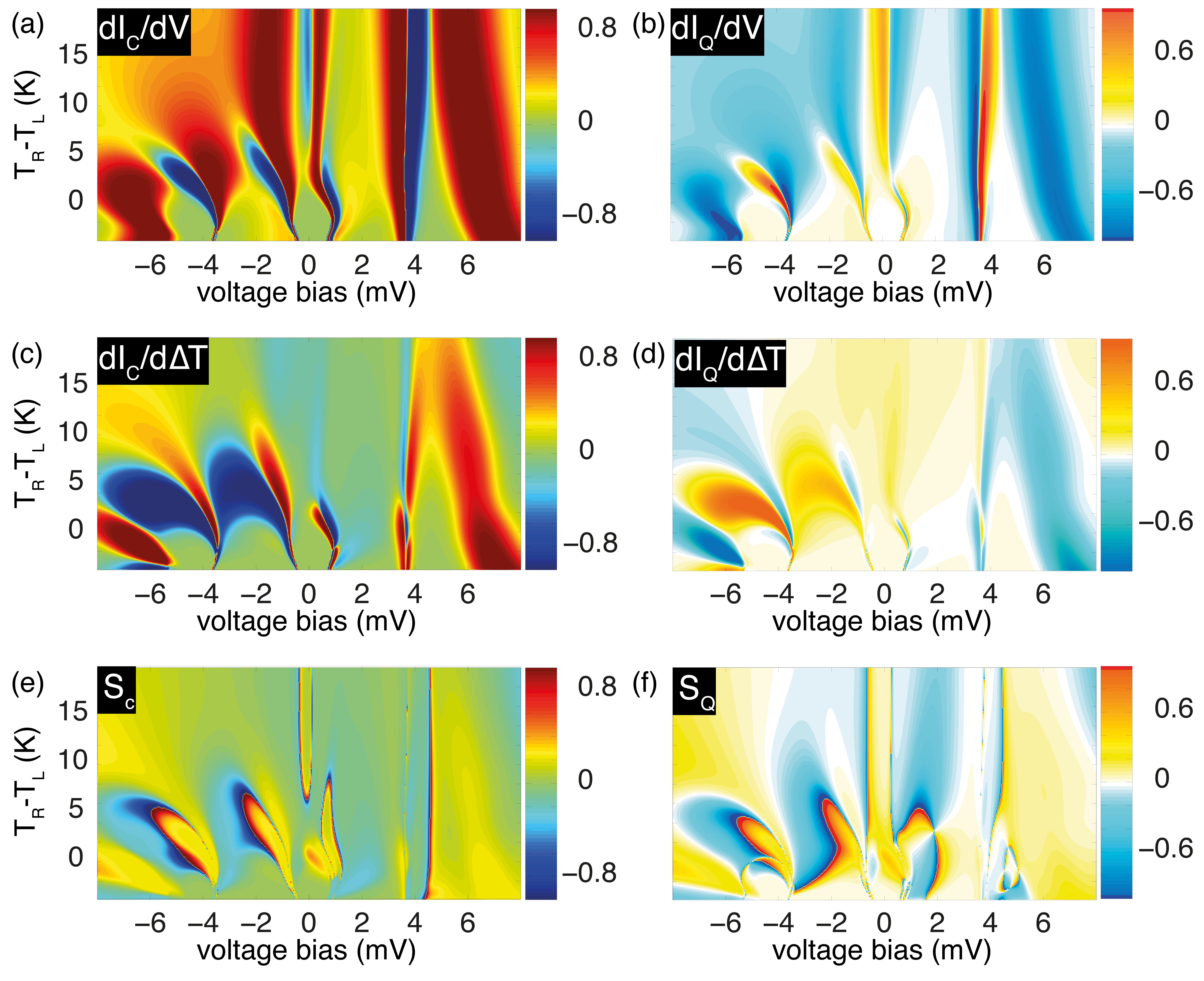}
\end{center}
\caption{(Color online)
(a) -- (d) Differential charge conductance (a), (c), and entropy production rate (b), (d), with respect to the voltage bias (a), (b), and temperature difference (c), (d).
(e), (f), Seebeck coefficients associated with the (e) charge, $S_c$, and (f) heat, $S_Q$, currents.
Here, $\dote{0}-\mu=1$, while other parameters are as in Figure \ref{fig-Fig1}.
All contours show $(1+\exp\{\sigma{\cal F}\})^{-1}$, where ${\cal F}$ is one of $\partial I_x/\partial V$, $\partial I_x/\partial\Delta T$, and $S_x$, whereas $\sigma$ is a scaling parameter.
}
\label{fig-conde1}
\end{figure}

Finally, we briefly discuss the differential conductances, both charge conductance and differential entropy production rate and with respect to both voltage bias and temperature difference. Hence, we calculate $\partial I_x/\partial V$ and $\partial I_x/\partial\Delta T$, where $x=c,Q$, see Figs. \ref{fig-conde0} (a) -- (d) and \ref{fig-conde1} (a) -- (d). In Figs. \ref{fig-conde0} and \ref{fig-conde1} we plot $1/(1+\exp\{\sigma\cal F\})$, where ${\cal F}$ is one of $\partial I_x/\partial V$, $\partial I_x/\partial\Delta T$, and $S_x$, in order to lower the values of the high amplitudes. First we notice that both currents have ranges with fairly rapid variations with the voltage bias and with the temperature difference. While these properties can be traced back to the corresponding variations of the exchange interaction, we can predict a few consequences of these features on the ratios of the differential conductance with respect to the temperature difference and the differential conductance with respect to the voltage bias, that is,
\begin{align}
S_c=&
	-\frac{\partial I_c/d\Delta T}{\partial I_c/dV}
	,
\end{align}
where the notation $S_c$ is used since the ratio coincides with the Seebeck coefficient in the limit $V\rightarrow0$, $\Delta T\rightarrow0$ \cite{Fransson2011}. In this sense the ratio $S_c$ is a non-equilibrium analogue of the Seebeck coefficient, however, we shall refrain from that nomenclature for sake of not causing confusion with the concepts.

Nontheless, as rapid variations in the charge current yield a large corresponding conductance and, oppositely, for slow variations lead to a small conductance, we would expect that $S_c$ typically is large in the regions with weak variations of the charge current. One is therefore led to think that $S_c$ might be large in the anti-ferromagnetic regime since the current is both small and slowly varying with the voltage bias in those regimes, see Figs. \ref{fig-conde0} (a) and \ref{fig-conde1} (a). In addition, within the anti-ferromagnetic regimes, $\partial I_c/\partial\Delta T$ varies rapidly, including possible sign changes, near zero temperature difference between the leads. This conjecture is fairly well corroborated in the plots of the $S_c$ shown in Figs. \ref{fig-conde0} (e) and \ref{fig-conde1} (e). In particular, it can be noticed in Figure \ref{fig-conde0} (e) that $S_c$ acquires large values in the anti-ferromagnetic regime (voltages roughly in the ranges $(-4,-2)$ and $(2,4)$ mV) for temperature differences between 0 K and 5 K. Except for these small regions of large $S_c$, it tends to be small in the remainder of the magnetically active regimes although not vanishing. In the case with a finite level off-set, Figure \ref{fig-conde1} (e), however, $S_c$ tends to be large in the regions where the anti-ferromagnetic coupling is destroyed by the increased temperature difference. There are clearly visible domains at negative voltages which can be correlated with the cross over between the anti-ferromagnetic and paramagnetic regimes. At positive voltages where the dimer remains magnetically active in a larger temperature range, $S_c$ is again small. We conclude that $S_c$ in both the ferromagnetic and anti-ferromagnetic regimes is small and typically becomes finite at the crossovers to the paramagnetic regime. Hence, the spin-spin interaction provides a way to not only control the charge current but also $S_c$ in the system.

In analogy with the definition of $S_c$ associated with the charge current, we can define a coefficient $S_Q$ for the entropy production rate through
\begin{align}
S_Q=&
	-\frac{\partial I_Q/d\Delta T}{\partial I_Q/dV}
	.
\end{align}
Although we cannot give this coefficient an as simple interpretation as with the thermopower, we nevertheless find it useful in the analysis of the influence of the magnetic configurations on the thermal properties. It can be seen in Figs. \ref{fig-conde0} (b) and \ref{fig-conde1} (b) that the differential entropy conductance, with respect to the voltage bias, has a non-trivial dependence on the voltage bias and temperature bias. Moreover, the dependence on the temperature difference is intriguing. We can, however, notice regarding the generic features of $S_Q$ that its more or less vanishing in the magnetically active regimes except in the crossover between the different regimes, where the entropy conductance varies slowly with the voltage bias but not necessarily with the temperature difference. The qualitative difference of the entropy conductance compared to the charge current in the ferromagnetic regime, leads to that $S_Q$ provides a complimentary piece of information about the system under investigation in addition to $S_c$.

\section{Conclusions}
\label{sec-summary}
In summary we have theoretically studied the transport properties of a dimer of paramagnetic molecules with localized spins embedded in a junction between metallic leads. In particular, we have addressed the charge and entropy conductance flowing through the junction. It is demonstrated that the indirect exchange interaction between the localized spins, which previously has been shown to depend on the voltage bias and temperature difference across the junction, acquires a strongly asymmetric voltage bias dependence under finite temperature difference between the leads. This property was subsequently is predicted to have implications on, for example, thermal rectification for both the charge and entropy conductance. Simultaneously, our calculations suggest that the temperature of the source electrode has a stronger influence on the properties of the indirect exchange than the drain. It was found, for instance, that while a voltage drop from the hotter to the colder reservoir tends to effectively destroy the tunable properties of the indirect exchange, these properties are stable under the opposite voltage polarity.

The transport properties of the dimer are intimately related to the indirect exchange, where the charge current is nearly blockaded for anti-ferromagnetic exchange, whereas it is finite for ferromagnetic exchange, and maximal whenever the exchange is negligible. These three regimes can be explained by different characteristics of the electronic density. In the anti-ferromagnetic regimes, the spin projections of the density is strongly localized to one molecule such that transport between the molecules is suppressed. When in the ferromagnetic regime, the electron density is partially delocalized which leads to an enhanced conductance, whereas the current can flow freely in the paramagnetic regime due to a completely delocalized density. The entropy conductance follows the same characteristics, as the charge current, in the anti-ferromagnetic and paramagnetic regimes. In the ferromagnetic regimes, however, the entropy conductance is strongly suppressed by a contribution which is large, and nearly cancels the regular entropy conductance contribution, near voltage biases corresponding to the energy of the resonant states in the molecular structure. By necessity, this resonant behavior occurs in the ferromagnetic regimes which leads to a strongly suppressed entropy conductance there as well.

We, further, demonstrated that the non-equilibrium thermopower, in general is finite at the cross-over between regimes of different indirect exchange associated with small differential conductance. Typically, this behavior can be summarized in that the thermopower is low within both the ferro- and anti-ferromagnetic regimes. Analogously, we introduced a thermal coefficient as the ratio between the differential entropy conductance with respect to the temperature difference and voltage bias. While some of its properties closely follow those ratios, additional features can be extracted, especially at the cross over between the ferro- and antiferromagnetic regimes. In this sense, this ratio provides a complementary sensitivity to the transport measurement which may show useful in the analysis of the internal properties of the system.

We remark that while our results presented in this article are quite qualitative, they are realistic from the following point of view. The results for the exchange interaction $J$ presented in the previous section, are obtained using the bare single electron Green functions for the molecular levels. This means that the back-action effect from the local spin moment is not included. In the calculations of the transport properties, on the other hand, the presence of the local spin moments are included, something which is crucial in order to investigate possible signatures in the transport data that originates from the spin configurations. 
As for the transport calculations we could have chosen to simply demonstrate how the charge and thermal transport depend on the nature of the exchange interaction, whether it is ferromagnetic (negative), anti-ferromagnetic (positive), or paramagnetic (zero). Given the approximation in which we calculate the local electronic Green function of the molecular dimer, we would obtain the transport characteristics that are presented in the Results. However, instead of making assumptions about the values of the exchange interaction, we use the values as calculated by the formulas provided in Exchange. In this way we incorporate the voltage bias and thermal gradient dependence of the exchange also on the transport properties.
While this approach certainly has its limitations, we notice that a more thorough study of the correspondence between the regimes with different spin couplings and the associated transport properties requires self-consistent calculations. Such calculations are, however, beyond the scope of the present investigation.

Considering the limitations of the method, we yet believe that our findings are realistic and relevant to existing molecular structures. The values of the local exchange interaction between localized spin and delocalized electrons can vary between 0.5 -- 20 meV \cite{Coronado2004,Chen2008}, which leaves a large window of tuning freedom with respect to couplings to the leads and HOMO/LUMO level off-set of the molecules. Moreover, since our predictions are stable with respect to differences in the local exchanges as well as the couplings to the leads, this also allows for flexibility in the design of potential experiments. 

The predictions discussed in this paper opens the possibility to design nanoscale structures, in particular using magnetic molecules, that have a strong sensitivity on the local spin states of the system, which can be measured through the charge and thermal transport characteristics. In ways, this suggests an alternative utilization of the spin degrees of freedom, compared to the conventionally implemented spintronics, in which external magnetic fields are absent. The absence of such fields, in turn, leads to that the spin-spin interactions are isotropic which implies a truly magnetically isotropic (paramagnetic) device. Experimental confirmation of our predictions are feasible using state-of-the-art technology.

\section{Acknowledgement}
We thank S. Borlenghi Garoia, A. Bouhon, K. Bj\"ornson, H. Hammar, D. Kuzmanovksi and J. Schmidt for fruitful discussions. Financial support from Colciencias (Colombian Administrative Department of Science, Technology and Innovation) and Vetenskapsr\aa det is acknowledged.


\begin{thebibliography}{99}

\bibitem{Huang1987} Huang, K. \emph{Statistical Mechanics}, {\bf 1987}, 2nd ed. (Wiley).
\bibitem{Nitzan2007} Nitzan, A. Chemistry. Molecules Take the Heat. \emph{Science} {\bf 2007}, \emph{317}, 759--760.
\bibitem{Pekola2015} Pekola, J. P. Towards Quantum Thermodynamics in Electronic Circuits. \emph{Nature Phys.} {\bf 2015}, \emph{11}, 118--123.
\bibitem{Reddy2007} Reddy, P.; Jang, S.-Y.; Segalman, R. A.; Majumdar, A. Thermoelectricity in molecular junctions. \emph{Science} {\bf 2007}, \emph{315}, 1568--1571.
\bibitem{Garrigues2016} Garrigues, A. R.; Wang, L.; del Barco, E.; Nijhuis, C. A. Electrostatic Control over Temperature-Dependent Tunnelling Across a Single-Molecule Junction. \emph{Nat. Comm.} {\bf 2016}, \emph{7}, 11595.
\bibitem{Osorio2010} Osorio, E. A.; Moth-Poulsen, K.; Van Der Zant, H. S. J.; Paaske, J.; Hedeg\aa rd, P.; Flensberg, K.; Bendix, J.; Bj\"ornholm, T. Electrical Manipulation of Spin States in a Single Electrostatically Gated Transition-Metal Complex. \emph{Nano Lett.} {\bf 2010}, \emph{10}, 105--110.
\bibitem{Parrondo2015} Parrondo, J. M. R.; Horowitz, J. M.; Sagawa, T. Thermodynamics of Information. \emph{Nat. Phys.} {\bf 2015}, \emph{11}, 131--139.
\bibitem{Esposito2015a} Esposito, M.; Ochoa, M. A.; Galperin, M. Quantum Thermodynamics: A Nonequilibrium Green's Function Approach. \emph{Phys. Rev. Lett.} {\bf 2015}, \emph{114}, 080602.
\bibitem{Shastry2016} Shastry, A.; Stafford, C. A. Temperature and Voltage Measurement in Quantum Systems Far From Equilibrium. \emph{Phys. Rev. B: Condens. Matter} {\bf 2016}, \emph{94}, 155433.
\bibitem{Dauxois2006} Dauxois, T.; Peyrard, M. \emph{Physics of Solitons} {\bf 2006} (Cambridge University Press, Cambridge).
\bibitem{Chiang1977} Chiang, C.K.; Fincher, C.R.; Park, Y.W.; Heeger, A.J.; Shirakawa, H.; Louis, E. J.; Gau, S. C.; MacDiarmid, A. G. Electrical Conductivity in Doped Polyacetylene. \emph{Phys. Rev. Lett.} {\bf 1977}, \emph{39}, 1098--1101.
\bibitem{Fincher1978} Fincher, C. J.; Peebles, D.; Heeger, A.; Druy, M.; Matsumura, Y.; MacDiarmid, A.; Shirakawa, H.; Ikeda, S. Anisotropic Optical Properties of Pure and Doped Polyacetylene. \emph{Sol. State Comm.} {\bf 1978}, \emph{27}, 489--494.
\bibitem{Schrieffer1980} Su, W. P.; Schrieffer, J. R.; Hegger, A. J. Soliton Excitations in Polyacetylene. \emph{Phys. Rev. B} {\bf 1980}, \emph{22}, 2099--2111.

\bibitem{PhysRevB.94.035420} Ochoa, M. A.; Bruch, A.; Nitzan, A. Energy distribution and local fluctuations in strongly coupled open quantum systems: The extended resonant level model, \emph{Phys. Rev. B: Condens. Matter Phys.} {\bf 2016} \emph{94}, 035420.

\bibitem{Rammer1986} Rammer, J.; Smith, H. Quantum Field-Theoretical Methods in Transport Theory of Metals. \emph{Rev. Mod. Phys.} {\bf 1986}, \emph{58}, 323--359.
\bibitem{Heeger1988} Heeger, A. J.; Kivelson, S.; Schrieffer, J. R.; Su, W. P. Solitons in Conducting Polymers. \emph{Rev. Mod. Phys.} {\bf 1988}, \emph{60}, 781--850.
\bibitem{Roth1989} Roth, S.; Bleier, H.; Pukacki, W. Charge Transport in Conducting Polymers. \emph{Faraday Discussions of the Chem. Soc.} {\bf 1989}, \emph{88}, 223--233.
\bibitem{Nakata1992} Nakata, M.; Taga, M.; Kise, H. Synthesis of Electrical Conductive Polypyrrole Films by Interphase Oxidative Polymerization -- Effects of Polymerization Temperature and Oxidizing Agents. \emph{Polym. Journ.} {\bf 1992}, \emph{24}, 437--441.
\bibitem{Meir1992} Meir, Y.; Wingreen, N. S. Landauer Formula for the Current Through an Interacting Electron Region. \emph{Phys. Rev. Lett.} {\bf 1992}, \emph{68}, 2512--2515.
\bibitem{Bao1996} Bao, Z.; Lovinger, A. J.; Dodabalapur, A. Organic Field-Effect Transistors with High Mobility Based on Copper Phthalocyanine. \emph{Appl. Phys. Lett.} {\bf 1996}, \emph{69}, 3066--3068.
\bibitem{Parthasarathy1998} Parthasarathy, G.; Burrows, P. E.; Khalfin, V.; Kozlov, V. G.; Forrest, S. R. A Metal-Free Cathode for Organic Semiconductor Devices. \emph{Appl. Phys. Lett.} {\bf 1988}, \emph{72}, 2138--2140.
\bibitem{Chen2002} Chen, G.; Shakouri, A. Heat Transfer in Nanostructures for Solid-State Energy Conversion. \emph{J. of Therm. Transf.} {\bf 2002}, \emph{124}, 242.
\bibitem{Chen2003} Chen, G.; Chen, G. Nanoscale Heat Transfer and Information Technology. \emph{Appl. Phys. Lett.} {\bf 2003}, \emph{29}, 1--3.
\bibitem{Haibo2009} Haibo, M.; Schollwock, U. Dynamical Simulations of Polaron Transport in Conjugated Polymers with the Inclusion of Electron-Electron Interactions. \emph{J. Phys. Chem. A} {\bf 2009}, \emph{113}, 1360--1367.
\bibitem{Haug2013} Haug, H. J.; Jauho, A.-P. \emph{Quantum Kinetics in Transport and Optics of Semiconductors}, {\bf 2008}, 2nd ed. (Springer, Berlin).

\bibitem{Giazotto2006} Giazotto, F.; Heikkil\"a, T.T.; Luukanen, A.; Savin, A.M.; Pekola, J. P. Opportunities for Mesoscopics in Thermometry and Refrigeration: Physics and Applications. \emph{Rev. Mod. Phys.} {\bf 2006}, \emph{78}, 217--274.
\bibitem{Dubi2011} Dubi, Y.; Di Ventra, M. Colloquium: Heat Flow and Thermoelectricity in Atomic and Molecular Junctions. \emph{Rev. Mod. Phys.} {\bf 2011}, \emph{83}, 131-155.
\bibitem{Liu2009} Liu, Y. S.; Chen, Y. C. Seebeck Coefficient of Thermoelectric Molecular Junctions: First-Principles Calculations. \emph{Phys. Rev. B: Condens. Matter Phys.} {\bf 2009}, \emph{79}, 193101.
\bibitem{Ludovico2014} Ludovico, M. F.; Lim, J. S.; Moskalets, M.; Arrachea, L.; S\'anchez, D. Dynamical Energy Transfer in ac-Driven Quantum Systems. \emph{Phys. Rev. B} {\bf 2014}, \emph{89}, 161306.
\bibitem{PhysRevB.92.235440} Esposito, M. Ochoa, M. A. and Galperin, M., Nature of Thermal in Strongly Coupled Open Quantum Systems. \emph{Phys. Rev. B: Condens. Matter Phys.}, {\bf 2015}, \emph{92}, 235440.
\bibitem{Dare2016} Dar\'e, A.M.; Lombardo, P. Time-Dependent Thermoelectric Transport for Nanoscale Thermal Machines. \emph{Phys. Rev. B} {\bf 2016}, \emph{93}, 035303.
\bibitem{Arrachea2014} Arrachea, L.; Bode, N.; von Oppen, F. Vibrational Cooling and Thermoelectric Response of Nanoelectromechanical Systems. \emph{Phys. Rev. B} {\bf 2014}, \emph{90}, 125450.
\bibitem{Zhou2015} Zhou, H.; Thingna, J.; H\"anggi, P.; Wang, J.-S.; Li, B. Boosting Thermoelectric Efficiency Using Time-Dependent Control. \emph{Sci. Rep.} {\bf 2015}, \emph{5}, 14870.
\bibitem{Segal2015a} Segal, D.; Agarwalla, B. K. Vibrational Heat Transport in Molecular Junctions. \emph{Ann. Rev. Phys. Chem.} {\bf 2016}, \emph{67}, 185--209.

\bibitem{Ye2016a} Ye, L.; Zheng, X.; Yan, Y.; Di Ventra, M. Thermodynamic Meaning of Local Temperature of Nonequilibrium Open Quantum Systems \emph{Phys. Rev. B} {\bf 2016}, \emph{94}, 245105.
\bibitem{White2013} White, A. J.; Peskin, U.; Galperin, M. Coherence in Charge and Energy Transfer in Molecular Junctions. \emph{Phys. Rev. B} {\bf 2013}, \emph{88}, 205424.
\bibitem{Eich2016} Eich, F. G.; Di Ventra, M.; Vignale, G. Temperature-Driven Transient Charge and Heat Currents in Nanoscale Conductors. \emph{Phys. Rev. B} {\bf 2016}, \emph{93}, 134309.
\bibitem{Wang2010} Wang, R. Q.; Sheng, L.; Shen, R.; Wang, B.; Xing, D. Y. Thermoelectric Effect in Single-Molecule-Magnet Junctions. \emph{Phys. Rev. Lett.} {\bf 2010}, \emph{105}, 057202.
\bibitem{Ramos-Andrade2016} Ramos-Andrade, J.P.; \'Avalos-Ovando, O.; Orellana, P.A.; Ulloa, S. E. Thermoelectric Transport Through Majorana Bound States and Violation of Wiedemann-Franz Law. \emph{Phys. Rev. B} {\bf 2016}, \emph{94}, 155436.
\bibitem{Kim2015} Kim, H. S.; Gibbs, Z. M.; Tang, Y.; Wang, H.; Snyder, G. J. Characterization of Lorenz Number with Seebeck Coefficient Measurement. \emph{APL Mater.} {\bf 2015}, \emph{3}, 041506.
\bibitem{Crepieux2011} Cr\'epieux, A.; \v{S}imkovic, F.; Cambon, B.; Michelini, F. Enhanced Thermopower Under a Time-Dependent Gate Voltage. \emph{Phys. Rev. B} {\bf 2011}, \emph{83}, 153417.
\bibitem{Bergfield2009} Bergfield, J. P.; Stafford, C. A. Thermoelectric Signatures of Coherent Transport in Single-Molecule Heterojunctions. \emph{Nano Lett.} {\bf 2009}, \emph{9}, 3072--3076.

\bibitem{Koole2015a} Koole, M.; Thijssen, J. M.; Valkenier, H.; Hummelen, J. C.; Van Der Zant, H. S. J. Electric-Field Control of Interfering Transport Pathways in a Single-Molecule Anthraquinone Transistor. \emph{Nano Lett.} {\bf 2015}, \emph{15}, 5569--5573.

\bibitem{Valkenier2014} Valkenier, H.; Gu\'edon, C.M.; Markussen, T.; Thygesen, K.S.; van der Molen, S. J.; Hummelen, J. C. Cross-Conjugation and Quantum Interference: A General Correlation? \emph{Phys. Chem. Chem. Phys.} {\bf 2014}, \emph{16}, 653--662.
\bibitem{Fracasso2011} Fracasso, D.; Valkenier, H.; Hummelen, J. C.; Solomon, G. C.; Chiechi, R. C. Evidence for Quantum Interference in Sams of Arylethynylene Thiolates in Tunneling Junctions with Eutectic Ga-In (EGaIn) Top-Contacts. \emph{J. Amer. Chem. Soc.} {\bf 2011}, \emph{133}, 9556--9563.
\bibitem{Aradhya2012} Aradhya, S. V.; Meisner, J. S.; Krikorian, M.; Ahn, S.; Parameswaran, R.; Steigerwald, M. L.; Nuckolls, C.; Venkataraman, L. Dissecting Contact Mechanics from Quantum Interference in Single-Molecule Junctions of Stilbene Derivatives. \emph{Nano Lett.} {\bf 2012}, \emph{12}, 1643--1647.
\bibitem{Guedon2012a} Gu\'edon, C.; Valkenier, H.; Markussen, T.; Thygesen, K.; Hummelen, J.; van der Molen, S. Observation of Quantum Interference in Molecular
Charge Transport. \emph{Nat. Nano.} {\bf 2012}, \emph{7}, 305--309.
\bibitem{Markussen2014} Markussen, T.; Thygesen, K. S. Temperature Effects on Quantum Interference in Molecular Junctions. \emph{Phys. Rev. B} {\bf 2014}, \emph{89}, 085420.
\bibitem{Bessis2016a} Bessis, C.; Rocca, M. L. D.; Barraud, C.; Martin, P.; Lacroix, J. C.; Markussen, T. Probing Electron-Phonon Excitations in Molecular Junctions by Quantum Interference. \emph{Sci. Rep.} {\bf 2016}, \emph{6}, 20899.

\bibitem{Darwish2012} Darwish, N.; D\'iez-P\'erez, I.; Da Silva, P.; Tao, N.; Gooding, J. J.; Paddon-Row, M. N. Observation of Electrochemically Controlled Quantum Interference in a Single Anthraquinone-Based Norbornylogous Bridge Molecule. \emph{Ang. Chem., Int. Ed.} {\bf 2012}, \emph{51}, 3203--3206.

\bibitem{Saygun2016a} Saygun, T.; Bylin, J.; Hammar, H.; Fransson, J. Voltage-Induced Switching Dynamics of a Coupled Spin Pair in a Molecular Junction. \emph{Nano Lett.} {\bf 2016}, \emph{16}, 2824-2829.
\bibitem{Diaz2012} D\'iazand, S.; N\'u\~{n}ez, \'A. S. Current-Induced Exchange Interactions and Effective Temperature in Localized Moment Systems. \emph{J. Phys: Condens. Matt.} {\bf 2012}, \emph{24}, 116001.
\bibitem{Fransson2014} Fransson, J.; Ren, J.; Zhu, J.-X. Electrical and Thermal Control of Magnetic Exchange Interactions. \emph{Phys. Rev. Lett.} {\bf 2014}, \emph{113}, 257201.
\bibitem{Fransson2008} Fransson, J.; Zhu, J. X. Spin Dynamics in a Tunnel Junction Between Ferromagnets. \emph{New J. Phys.} {\bf 2008}, \emph{10}, 013017.
\bibitem{Hammar2016} Hammar, H.; Fransson, J. Time-Dependent Spin and Transport Properties of a Single Molecule Magnet in a Tunnel Junction. \emph{Phys. Rev. B.} {\bf 2016}, \emph{94}, 054311.
\bibitem{Zimbovskaya2013} Zimbovskaya, N. A. \emph{Transport Properties of Molecular Junctions} {\bf 2013}, Vol. 254 (Springer, New York).
\bibitem{Moskalets2016} Moskalets, M.; Haack, G. Heat and Charge Transport Measurements to Access Single-Electron Quantum Characteristics. \emph{phys. stat. sol.} {\bf 2016}, \emph{254}, 1600616.
\bibitem{Gaury2014} Gaury, B.; Weston, J.; Santin, M.; Houzet, M.; Groth, C.; Waintal, X. Numerical Simulations of Time-Resolved Quantum Electronics. \emph{Phys. Rep.} {\bf 2014}, \emph{534}, 1--37.
\bibitem{Zimbovskaya2016} Zimbovskaya, N. A. Seebeck Effect in Molecular Junctions. \emph{J. Phys: Condens. Matt.} {\bf 2016}, \emph{28}, 183002.
\bibitem{Simmonds2012} Simmonds, R. W. Thermal Physics: Quantum Interference Heats Up. \emph{Nature} {\bf 2012}, \emph{492}, 358--359.
\bibitem{Heinrich2013} Heinrich, B. W.; Braun, L.; Pascual, J. I.; Franke, K. J. Protection of Excited Spin States by a Superconducting Energy Gap. \emph{Nat. Phys.} {\bf 2013}, \emph{9}, 765--768.
\bibitem{Heinrich2013a} Heinrich, B. W.; Ahmadi, G.; M\"uller, V. L.; Braun, L.; Pascual, J. I.; Franke, K. J. Change of the Magnetic Coupling of a Metal-Organic Complex with the Substrate by a Stepwise Ligand Reaction. \emph{Nano Lett.} {\bf 2013}, \emph{13}, 4840--4843.
\bibitem{Heinrich2015} Heinrich, B. W.; Braun, L.; Pascual, J. I.; Franke, K. J. Tuning the Magnetic Anisotropy of Single Molecules. \emph{Nano Lett.} {\bf 2015}, \emph{15}, 4024--4028.
\bibitem{Brumboiu2016} Brumboiu, I. E.; Haldar, S.; L\"uder, J.; Eriksson, O.; Herper, H. C.; Brena, B.; Sanyal, B. Influence of Electron Correlation on the Electronic Structure and Magnetism of Transition-Metal Phthalocyanines. \emph{J. Chem. Theor. Comp.} {\bf 2016}, \emph{12}, 1772--1785.
\bibitem{Urdampilleta2011} Urdampilleta, M.; Klyatskaya, S.; Cleuziou, J.-P.; Ruben, M.; Wernsdorfer, W. Supramolecular Spin Valves. \emph{Nat. Mater.} {\bf 2011}, \emph{10}, 502--506.
\bibitem{Vincent2012} Vincent, R.; Klyatskaya, S.; Ruben, M.; Wernsdorfer, W.; Balestro, F. Electronic Read-Out of a Single Nuclear Spin Using a Molecular Spin Transistor. \emph{Nature} {\bf 2012}, \emph{488}, 357--360.
\bibitem{Liu2014} Liu, Z. F.; Wei, S.; Yoon, H.; Adak, O.; Ponce, I.; Jiang, Y.; Jang, W. D.; Campos, L. M.; Venkataraman, L.; Neaton, J. B. Control of Single-Molecule Junction Conductance of Porphyrins via a Transition-Metal Center. \emph{Nano Lett.} {\bf 2014}, \emph{14}, 5365--5370.
\bibitem{Fransson2011} Fransson, J.; Galperin, M. Spin Seebeck Coefficient of a Molecular Spin Pump. \emph{Phys. Chem. Chem. Phys.} {\bf 2011}, \emph{13}, 14350-14357.
\bibitem{Misiorny2013} Misiorny, M.; Hell, M.; Wegewijs, M. R. Spintronic Magnetic Anisotropy. \emph{Nat. Phys.} {\bf 2013}, \emph{9}, 801--805.
\bibitem{Galperin2007} Galperin, M.; Nitzan, A.; Ratner, M. A. Heat Conduction in Molecular Transport Junctions. \emph{Phys. Rev. B.} {\bf 2007}, \emph{75}, 155312.
\bibitem{Jauho1994} Jauho, A. P.; Wingreen, N. S.; Meir, Y. Time-dependent Transport in Interacting and Noninteracting Resonant-Tunneling Systems. \emph{Phys. Rev. B.} {\bf 1994}, \emph{50}, 5528--5544.

\bibitem{Coronado2004} Coronado, E.; Day, P. Magnetic Molecular Conductors. \emph{Chem. Rev.} {\bf 2004}, \emph{104}, 5419--5448.
\bibitem{Chen2008} Chen, X.; Fu, Y.-S.; Ji, S.-H.; Zhang, T.; Cheng, P.; Ma, X.-C.; Zou, X.-L.; Duan, W.-H.; Jia, J.-F.; Xue, Q.-K. Probing Superexchange Interaction in Molecular Magnets by Spin-Flip Spectroscopy and Microscopy. \emph{Phys. Rev. Lett.} {\bf 2008}, \emph{101}, 197208.

\bibitem{Chudnovskiy2008} Chudnovskiy, A.~L.; Swiebodzinski, J.; Kamenev, A. Spin-Torque Shot Noise in Magnetic Tunnel Junctions. \emph{Phys. Rev. Lett.} {\bf 2008}, \emph{101}, 066601.
\bibitem{Ludwig2017} Ludwig, T.; Burmistrov, I.~S.; Gefen, Y.; Shnirman, A. Strong Nonequilibrium Effects in Spin-Torque Systems. \emph{Phys. Rev. B} {\bf 2017}, \emph{95}, 075425.
\bibitem{arXiv:1702.02820} Hammar, H.; Fransson, J. Transient Spin Dynamics in a Single-Molecule Magnet. \emph{Phys. Rev. B} {\bf 2017}, \emph{96}, 214401.


\end{thebibliography}
\end{document}